\documentclass{aa}
\usepackage{graphicx}

\begin{document}

\title{ISOCAM view of the starburst galaxies
       \object{M\,82}, \object{NGC\,253}, and \object{NGC\,1808}
     \thanks{Based on observations with {\em ISO\/}, an ESA project
               with instruments funded by ESA member states (especially
               the PI countries: France, Germany, the Netherlands, and
               the United Kingdom), and with participation of ISAS and NASA.}
}


\author{N. M. F\"orster Schreiber\inst{1}\fnmsep\thanks{\emph{Present address:}
          Leiden Observatory, PO Box 9513, 2300 RA Leiden, The Netherlands}
    \and M. Sauvage\inst{1}
    \and V. Charmandaris\inst{2, 3}
    \and O. Laurent\inst{1, 4}
    \and P. Gallais\inst{1}
    \and I. F. Mirabel\inst{1, 5}
    \and L. Vigroux\inst{1}}

\institute{CEA/DSM/DAPNIA/Service d'Astrophysique, C. E. Saclay,
           F-91191 Gif sur Yvette CEDEX, France 
           \and
           Astronomy Department, Cornell University, 
           Ithaca, NY 14853, USA
	   \and 
	   Chercheur Associ\'e, Observatoire de Paris, LERMA,
	   61 Ave. de l'Observatoire, F-75014 Paris, France
	   \and
           Max-Planck-Institute f\"ur extraterrestrische Physik,
           Postfach 1312, D-85741 Garching, Germany 
	   \and
           Instituto de Astronom\'{\i}a y F\'{\i}sica del Espacio, 
           cc 67, suc 28, 1428 Buenos Aires, Argentina}

\offprints{N. M. F\"orster Schreiber,
  \email{forster@strw.leidenuniv.nl}}

\titlerunning{ISOCAM view of \object{M\,82}, \object{NGC\,253},
              and \object{NGC\,1808}}
\authorrunning{F\"orster Schreiber et al.}

\date{Received 22 August 2002; accepted 15 November 2002}

\abstract{
We present results of mid-infrared $\lambda = 5.0 - 16.5~\mathrm{\mu m}$
spectrophotometric imaging of the starburst galaxies \object{M\,82},
\object{NGC\,253}, and \object{NGC\,1808} from the ISOCAM instrument
on board the {\em Infrared Space Observatory\/}.  The mid-infrared spectra
of the three galaxies are very similar in terms of features present.  
The $\lambda \ga 11~\mathrm{\mu m}$ continuum attributed to very small
dust grains (VSGs) exhibits a large spread in intensity relative to the
short-wavelength emission.  We find that the 15$\,\mu$m dust continuum flux
density correlates well with the fine-structure [\ion{Ar}{ii}] 6.99$\,\mu$m
line flux and thus provides a good quantitative indicator of the level of
star formation activity.  By contrast, the $\lambda = 5 - 11~\mathrm{\mu m}$ 
region dominated by emission from polycyclic aromatic hydrocarbons (PAHs) has
a nearly invariant shape.  Variations in the relative intensities of the PAH
features are nevertheless observed, at the $20\% - 100\%$ level.  We illustrate
extinction effects on the shape of the mid-infrared spectrum of obscured
starbursts, emphasizing the differences depending on the applicable extinction
law and the consequences for the interpretation of PAH ratios and extinction
estimates.  The relative spatial distributions of the PAH, VSG, and 
[\ion{Ar}{ii}] 6.99$\,\mu$m emission between the three galaxies exhibit
remarkable differences.  The $\la 1~\mathrm{kpc}$ size of the mid-infrared
source is much smaller than the optical extent of our sample galaxies and
$70\% - 100\%$ of the {\em IRAS\/} 12$\,\mu$m flux is recovered within the
ISOCAM $\leq 1.5~\mathrm{arcmin^{2}}$ field of view, indicating that the
nuclear starburst dominates the total mid-infrared emission while diffuse
light from quiescent disk star formation contributes little.
}

\maketitle

\section{Introduction}  \label{Sect-intro}

As hosts of vigorous episodes of star formation activity, starburst galaxies 
in the local Universe offer a unique opportunity to study the star formation
process in extreme environments reminiscent of the conditions presumably
prevailing in primordial galaxies.  Studies in recent years have underscored
the prominent role of starbursts in galaxy formation and evolution as well as
their significant contribution to the extragalactic background, and to the
chemical enrichment and heating of the intergalactic medium up to the highest
redshifts (e.g. Steidel et al.~\cite{Ste96}; Puget et al.~\cite{Pug96};
Smail et al.~\cite{Sma97}; Franx et al.~\cite{Fra97};
Pettini et al.~\cite{Pet98}; Elbaz et al.~\cite{Elb02}).
Understanding the starburst phenomenon is thus a key issue of local and
cosmological relevance.

Nuclear starburst galaxies generally emit the bulk of their luminosity in the
infrared, primarily as reprocessed radiation by interstellar dust grains
heated by important populations of hot massive stars.  In addition, a
substantial fraction -- if not most -- of the star formation in starburst
systems is heavily obscured at optical and ultraviolet wavelengths by large
amounts of dust.  This has been particularly dramatically illustrated by
mid-infrared imaging of the colliding pair \object{NGC\,4038/4039}
(Mirabel et al.~\cite{Mir98}).  Studies at infrared wavelengths are thus
crucial in defining the properties of starburst galaxies and the interplay
between the primary energy sources and the reprocessing material.  The
growing evidence for the presence and significant role of dust in distant
galaxies further emphasizes the need for infrared investigations of nearby
dust-rich templates (e.g. Rowan-Robinson \cite{Row01};
Blain et al.~\cite{Bla02}; and references therein).
However, except for a limited sample, the detailed spatial and spectral energy
distributions of starburst galaxies in this regime are still poorly known.
Progress has been hindered by instrumental limitations, and by the difficulties
inherent to ground-based and air-borne observations at these wavelengths.

The {\em Infrared Space Observatory\/} ({\em ISO\/};
Kessler et al.~\cite{Kes96}) has revolutionized
the field of infrared astronomy (see the reviews by 
Cesarsky \& Sauvage \cite{Ces99}; Genzel \& Cesarsky \cite{Gen00}),
with unprecedented capabilities compared to its predecessor the
{\em Infrared Astronomical Satellite} ({\em IRAS}\/).  In the mid-infrared
(MIR, $\lambda = 5.0 - 16.5~\mathrm{\mu m}$), the camera ISOCAM
(Cesarsky et al.~\cite{Ces96}) provided an improvement in sensitivity and
spatial resolution by factors of $\sim 1000$ and $\sim 60$, respectively.
In addition to broad- and narrow-band filters, ISOCAM featured continuously
variable filters (CVFs) allowing arcsecond-scale spectrophotometric imaging 
with full wavelength coverage at a resolution of 
$R \equiv \lambda/\Delta\lambda \sim 40$.  The MIR regime contains a variety
of key features tracing different components of the interstellar medium (ISM)
and is subject to relatively little extinction by dust 
($A_\mathrm{5 - 17\,\mu m} < 0.1\,A_{V}$).
Spectrophotometric observations permit studies of the most prominent features
to characterize the spatial distribution and infer the properties of the ISM
and exciting sources, deep into obscured star-forming regions.

In this paper, we present results of ISOCAM CVF observations
of \object{M\,82}, \object{NGC\,253}, and \object{NGC\,1808}.
\object{M\,82} and \object{NGC\,253} are two of the nearest
and brightest starburst galaxies (at distances of $D = 3.3$
and 2.5~Mpc, respectively, $1^{\prime\prime} = 16$ and 12~pc;
Freedman \& Madore \cite{Fre88}; Davidge \& Pritchet \cite{Dav90}).
Both have been extensively studied in all accessible spectral regions
and have often been considered as prototypical objects for the
starburst phenomenon (see Telesco \cite{Tel88}, Rieke et al.~\cite{Rie93},
and Engelbracht et al.~\cite{Eng98} for reviews).
\object{NGC\,1808} is a more distant system ($D = 10.9~\mathrm{Mpc}$
for $H_{0} = 75~\mathrm{km\,s^{-1}\,Mpc^{-1}}$, 
$1^{\prime\prime} = 53~\mathrm{pc}$; Sandage \& Tammann \cite{San87})
which closely resembles \object{M\,82} and \object{NGC\,253} in its global
properties (e.g. Dahlem et al.~\cite{Dah90}; Junkes et al.~\cite{Jun95}).
In all three galaxies, starburst activity takes place within the central
$\sim 0.1 - 1~\mathrm{kpc}$ regions which are severely obscured at optical
and ultraviolet wavelengths.  The large extinction levels likely result in
part from high disk inclination angles of $80^{\circ}$, $78^{\circ}$, 
and $57^{\circ}$ for \object{M\,82}, \object{NGC\,253}, and \object{NGC\,1808}
(G\"otz et al.~\cite{Got90}; Pence \cite{Pen81}; Reif et al.~\cite{Rei82}).

At comparable infrared luminosities\footnote{as computed from the {\em IRAS\/} 
fluxes following the prescription of Sanders \& Mirabel (\cite{San96}).}
$L_\mathrm{IR} \equiv L_\mathrm{8-1000\,\mu m}$ of
$3.7 \times 10^{10}$, $1.4 \times 10^{10}$, and
$3.4 \times 10^{10}~\mathrm{L_{\odot}}$ for \object{M\,82},
\object{NGC\,253}, and \object{NGC\,1808},
these galaxies also share key morphological features including a
kiloparsec-scale stellar bar, large amounts of molecular gas in the central
$\sim 1~\mathrm{kpc}$ predominantly concentrated in circumnuclear ring-,
spiral-, or bar-like structures, populations of compact and luminous
``super star clusters'' as well as of compact non-thermal radio sources
associated with supernova remnants, and a large-scale outflowing starburst
wind (F\"orster Schreiber et al.~\cite{FS01}; Engelbracht et al.~\cite{Eng98};
Tacconi-Garman et al.~\cite{Tac96}; and references therein).
The triggering of starburst activity in \object{M\,82} is generally
attributed to gravitational interaction as evidenced by the extended
filamentary tidal features in \ion{H}{i} gas distribution threading
\object{M\,82} and its neighbours \object{M\,81} and \object{NGC\,3077}
or, alternatively, to the stellar bar which may itself have been induced
by the interaction (e.g. Yun et al.~\cite{Yun93, Yun94}).
In \object{NGC\,253}, the bar itself appears to be the primary mechanism
responsible for the onset of starburst activity as there are no obvious
signatures of interaction with a nearby companion, although a past merger
or accretion event involving a small galaxy has been suggested based on
kinematical evidence (e.g. Anantharamaiah \& Goss \cite{Ana96};
B\"oker et al.~\cite{Bok98}).
Tidal interaction with \object{NGC\,1792} has been proposed
for \object{NGC\,1808} based on circumstantial evidence
(e.g. Dahlem et al.~\cite{Dah90}; Koribalski et al.~\cite{Kor93}).
The recent high sensitivity \ion{H}{i} observations by Dahlem et al.
(\cite{Dah01}) do not indicate the presence of any tidal feature,
dwarf galaxy or other debris that could support a close passage in
the past, leaving the stellar bar as more probable trigger or a
possible merger/accretion event.

In view of their similar nature, we focus on the comparison of the
spatial and spectral properties of \object{M\,82}, \object{NGC\,253},
and \object{NGC\,1808} at MIR wavelengths.
Though small, this sample also allows us to probe the transition regime
between normal spiral and irregular galaxies and the more extreme luminous
and ultraluminous infrared galaxies (LIRGs and ULIRGs for which
$L_\mathrm{IR} > 10^{11}$ and $> 10^{12}~\mathrm{L_{\odot}}$, respectively).
Section~\ref{Sect-obs} describes the observations and data reduction 
procedure, and Sect.~\ref{Sect-res} presents the results.
Section~\ref{Sect-Spat_distr} discusses the origin and spatial distibution
of the continuum and emission features.
Section~\ref{Sect-Spec_distr} addresses the issues of extinction effects,
variations in spectral properties, and indicators of star formation activity.
Section~\ref{Sect-conclu} summarizes the paper.

\section{Observations and data reduction}  \label{Sect-obs}

\begin{figure*}[!ht]
\centering
\resizebox{1.00\hsize}{!}{\includegraphics{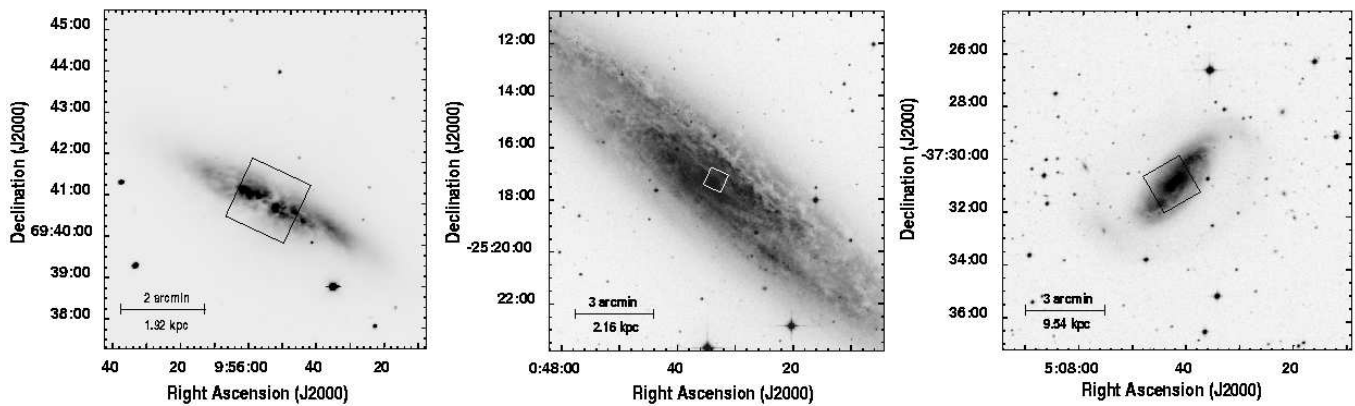}}
\caption
{
The ISOCAM CVF field of view, indicated by a square, is superimposed on
publicly available images in the $V$ band for M\,82 (left), and $R$ band
for NGC\,253 and NGC\,1808 (middle and right).
}
\label{fig-FOV}
\end{figure*}

\begin{table*}
\caption[]{Log of the observations  \label{tab-obs_log}}
\setlength{\tabcolsep}{0.35cm}
\begin{center}
\begin{tabular}{llllllll}
\hline\hline
Source & Date & Revolution & $\alpha_{2000}$ & $\delta_{2000}$ & 
P.A.\,$^{a}$ & Pixel scale & $t_\mathrm{int}$\,$^{b}$ \\
   &  &  & ($\mathrm{{ }^{h}:{ }^{m}:{ }^{s}}$) & 
(${ }^{\circ}:{ }^{\prime}:{ }^{\prime\prime}$) & 
(degree) & (arcsec) & (s) \\
\hline
M\,82     & 1996 Mar 19 & 123 & 09:55:51.8 & $+$69:40:45.8 & 212 & 3.0 & 25.2 \\
NGC\,253  & 1997 Jan 03 & 414 & 00:47:32.9 & $-$25:17:18.3 & 290 & 1.5 & 25.2 \\
NGC\,1808 & 1997 Oct 12 & 697 & 05:07:42.3 & $-$37:30:46.1 & 64 & 3.0 & 25.2 \\
\hline
\end{tabular}
\begin{list}{}{}
\item[$^{a}$] Position angle of the vertical axis of the ISOCAM array relative
              to the north and increasing counterclockwise.
\item[$^{b}$] Total integration time per wavelength channel.
\end{list}
\end{center}
\end{table*}

We observed \object{M\,82}, \object{NGC\,253}, and \object{NGC\,1808}
with ISOCAM (Cesarsky et al.~\cite{Ces96}) on board {\em ISO\/} 
(Kessler et al.~\cite{Kes96}) as part of the guaranteed time
program CAMACTIV (P.I. I. F. Mirabel).  We used the CVFs to cover the
entire $\lambda = 5.0 - 16.5~\mathrm{\mu m}$ range with a resolution of
$R \equiv \lambda/\Delta\lambda \approx 35$ to 45 and increments
$\delta\lambda = 0.06$ to 0.1$\,\mu$m from short to long wavelengths.
For all three sources, twelve single-frame exposures of 2.1~s were
recorded per wavelength channel, for total integration times of about
$\mathrm{1^{h} 05^{m}}$.
We selected the $3^{\prime\prime}\,\mathrm{pixel^{-1}}$
scale for \object{M\,82} and \object{NGC\,1808}, and the
$1.5^{\prime\prime}\,\mathrm{pixel^{-1}}$ scale for \object{NGC\,253};
total fields of view of $96^{\prime\prime} \times 96^{\prime\prime}$ and 
$48^{\prime\prime} \times 48^{\prime\prime}$ are thus covered by the
$32 \times 32$ pixels detector array.
Table~\ref{tab-obs_log} presents the details of the observations.
The areas observed with ISOCAM cover almost entirely the MIR sources
in each galaxy (Sect.~\ref{Sect-Spat_distr}) and are indicated on
optical images in Fig.~\ref{fig-FOV}.  This emphasizes the small
MIR source sizes compared to the optical extent of the galaxies and
that most of the MIR emission originates in their nuclear regions.

We reduced the data using the CAM Interactive Analysis package
(CIA)\footnote{CIA is a joint development by the ESA Astrophysics 
Division and the ISOCAM consortium led by the ISOCAM P.I. C. Cesarsky, 
Direction des Sciences de la Mati\`ere, C.E.A. Saclay, France.}
following the procedures described in the ISOCAM User's Manual
(Delaney \cite{Del97}).
We first subtracted the dark current using a model of the secular evolution
of ISOCAM's dark current (Biviano et al.~\cite{Biv97}), removed
cosmic ray hits by applying a multiresolution median filtering technique
(Starck et al.~\cite{Sta99}), and corrected for transient detector memory 
effects using the ``Fouks-Schubert'' algorithm 
(Coulais \& Abergel \cite{Cou00}).
We then combined the individual exposures for each wavelength channel
with the help of a shift-and-add algorithm accounting for the satellite 
jitter motions (typical amplitude of $\approx 1^{\prime\prime}$)
and performed absolute flux calibration based on the sensitivity
calibration files provided in CIA version 3.0.

For proper comparison of fluxes and maps obtained at different wavelengths,
we smoothed the CVF channels to a common angular resolution and
registered them appropriately.  The point spread function (PSF) varies 
throuhgout the CVF scans, with theoretical full-width at half-maximum 
$\mathrm{FWHM} = 3.1^{\prime\prime} - 5.6^{\prime\prime}$ for the
$3^{\prime\prime}\,\mathrm{pixel^{-1}}$ scale and
$2.0^{\prime\prime} - 5.2^{\prime\prime}$ for the
$1.5^{\prime\prime}\,\mathrm{pixel^{-1}}$ scale,
over the range $5.5 - 16.5~\mathrm{\mu m}$.
We convolved the data to a final resolution of 5.6\arcsec\ for
\object{M\,82} and \object{NGC\,1808}, and 5.2\arcsec\ for \object{NGC\,253},
corresponding to a linear resolution of 90, 297, and 62~pc, respectively.
We computed the transfer functions (close to Gaussians in profile)
using the library of theoretical PSFs for ISOCAM and a CLEAN algorithm.
The small-scale structure is thus smeared out in the higher
resolution channels but no important spatial feature is lost.

The source position varies slightly with wavelength due to telescope
motions and changes in optical path, with the largest offset occurring
near 9.2$\,\mu$m between the short and long wavelength segments of the CVFs.
Since no unresolved point source is detected in the data, we determined
the position offsets from cross-correlation over the entire field of view,
taking as references images integrated over various wavelength intervals.
Effects due to the different spatial distributions of
emission features were easily identified and ignored in the
determination of the instrumental shifts.  We fitted straight lines
to the cross-correlation offsets for each CVF segment.  For the
$3^{\prime\prime}\,\mathrm{pixel^{-1}}$ scale, the horizontal, or $x$-axis
offsets relative to 9.2$\,\mu$m in the short and long wavelength segments 
did not exceed 0.3~pixel and 0.5~pixel, respectively, with a jump of 
0.7~pixel at the transition wavelength.  Shifts along the $y$-axis were
negligible throughout the entire wavelength range ($< 0.2~\mathrm{pixel}$).
For the $1.5^{\prime\prime}\,\mathrm{pixel^{-1}}$ scale, both $x$- and 
$y$-axis offsets were $\leq 0.3~\mathrm{pixel}$ within each CVF segment,
with jumps at 9.2$\,\mu$m of 0.2~pixel only.

We did not attempt to correct for instrumental flat-field, straylight, 
and ghost images generated by multiple reflections between the detector
and filters since appropriate calibration files for the CVF mode were not
available.  Based on theoretical models for point-like and extended sources
as well as observations of stars and of the zodiacal light
(Biviano et al.~\cite{Biv98a, Biv98b}; Okumura \cite{Oku00}),
the combined flat-field, straylight, and ghost effects
integrated over the entire detector array vary between
$\approx 35\%$ and 10\% of the source flux from 5 to 16$\,\mu$m,
and exhibit qualitatively similar spatial structures at different
wavelengths.  The r.m.s. noise estimated from existing flat-field
images obtained with ISOCAM's broad-band filters is $\sim 10\%$
for both pixel scales (Biviano et al.~\cite{Biv98b}).

Based on available calibration accuracy reports 
(Blommaert \& Cesarsky \cite{Blo98} and references therein),
the systematic errors of the data processing and flux calibration
described above sum up to total uncertainties on the absolute
photometry of $\approx 50\%$.  For our analysis, we will consider
formal effective uncertainties on the {\em relative} fluxes estimated
as follows.  We measured the dispersion around the mean flux among
the individual reduced exposures for each wavelength channel at each
detector pixel, included an additional 10\% r.m.s. noise for intrinsic
pixel-to-pixel variations in sensitivity, and accounted for uncertainties
in relative spectral calibration.  For \object{M\,82} and \object{NGC\,253},
the latter were evaluated from the comparison with data from the {\em ISO\/}
Short Wavelength Spectrometer (SWS; de Graauw et al.~\cite{deG96})
in Sect.~\ref{Sect-res}, amounting to 5\% and 11\%, respectively.
For \object{NGC\,1808}, very low flux levels and high noise around 
10$\,\mu$m, below 6$\,\mu$m,
and outside of the main emission region led to poor correction
for transient memory effects because of the difficulty in determining the
initial and stabilized fluxes and of the bad behaviour of the algorithm 
wherever fluxes become negative.  In this case, we estimated the errors
in relative spectral calibration by comparing results when applying the 
transient correction to the data as they were and to the data where all
values lower than 0.01 were set to 0.01, adding a conservative 10\%. 
The median and average differences are of 16\% and a factor of 2.
Experimentation showed that the choice of the threshold is of little
consequence as long as it is sufficiently smaller than the signal where
the transient correction is satisfactory.  

Overall, the formal effective uncertainties per pixel and wavelength
channel within the brighter emission regions of interest (defined by
the ``starburst cores'' in Sect.~\ref{Sect-res}) have median values
of 18\%, 29\%, and 45\% for \object{M\,82}, \object{NGC\,253}, and
\object{NGC\,1808}.  Computed over the
entire ISOCAM field of view, the medians increase to 32\%, 88\%, and
a factor of 2.6 as a result of the higher noise towards the edges of
the array where little flux is detected.

\section{Results}  \label{Sect-res}

\subsection{Mid-infrared spectra and feature identification}  \label{Sub-spec}

We extracted from the reduced ISOCAM data cubes spectra integrated over
various regions: a ``starburst core'' centered on each galaxy's
nucleus\footnote{as defined by the peak of the near-infrared 2$\,\mu$m
emission; see table~\ref{tab-apertures} for coordinates and references.}
and covering most of the emission source, and a smaller region at the
position of the peak observed for the integrated MIR emission.
For \object{M\,82} and \object{NGC\,253}, we further included a
``disk'' corresponding to an annulus outside of the starburst core
and the ISOCAM field of view.  Due to the higher noise level
in the \object{NGC\,1808} data, only the starburst core
and MIR peak spectra are of sufficient quality for analysis.
The disk, core, and MIR peak regions were chosen to sample
relatively quiescent to intense star-forming activity as traced by 
e.g. the 15$\,\mu$m continuum emission (see Sects.~\ref{Sub-images} 
and \ref{Sub-cont}).  The starburst cores correspond to the apertures
generally used in previous studies when referring to the global
properties of the starburst in each galaxy.

\begin{figure}[!ht]
\centering
\resizebox{\hsize}{!}
          {\includegraphics[bb=17 80 340 680,clip]{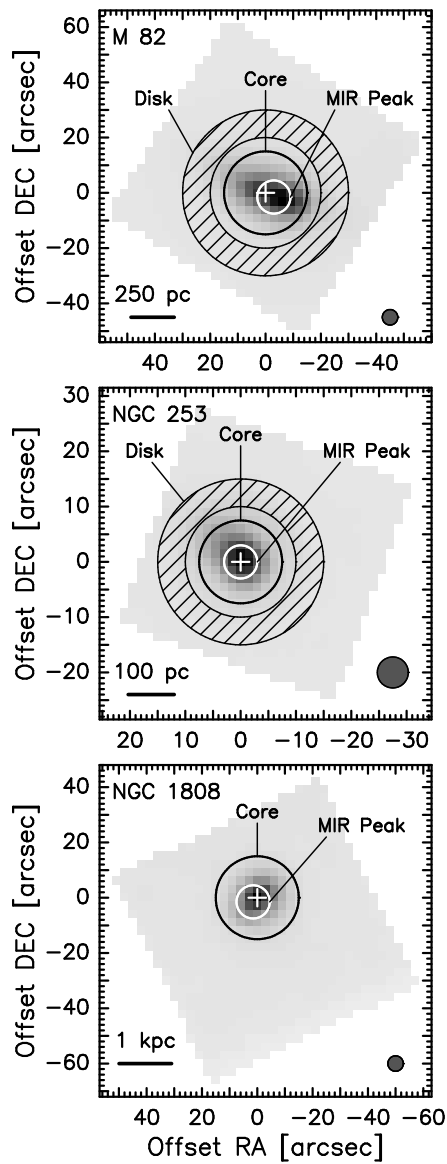}}
\caption
{
Selected regions in M\,82 (top), NGC\,253 (middle), and NGC\,1808
(bottom), shown on $\lambda = 14.8 - 15.2~\mathrm{\mu m}$ continuum maps.
The starburst core and MIR peak regions are enclosed within
the black and white circles, respectively.
The disk regions for M\,82 and NGC\,253 correspond to the hatched areas.
The parameters for all regions are given in table~\ref{tab-apertures}.
The horizontal bar and the filled circle at the bottom of each panel
indicate the physical linear scale and the FWHM of the PSF.
}
\label{fig-selreg}
\end{figure}

\begin{figure*}[!ht]
\centering
\resizebox{0.85\hsize}{!}{\includegraphics{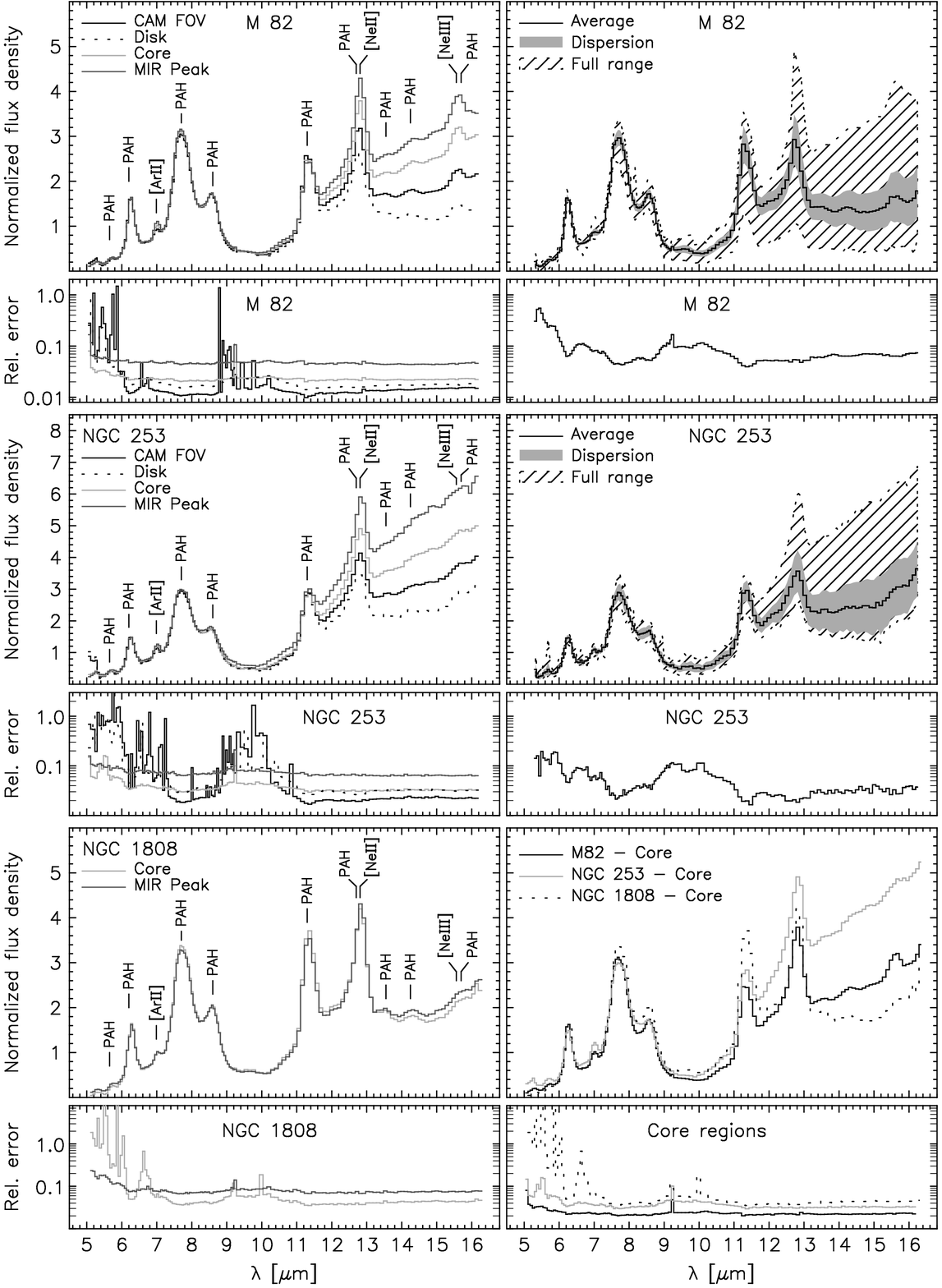}}
\caption
{
Mid-infrared spectra of M\,82, NGC\,253, and NGC\,1808.
All spectra are normalized to unit total flux density between 
6.0 and $6.6~\mathrm{\mu m}$.
The three panel pairs to the left show the spectra and relative
formal effective uncertainties for selected regions:
the ISOCAM field of view, the disk region, the starburst core, and
the MIR peak (see labels in each plot); the apertures used,
normalizing factors, and typical uncertainties are given in 
table~\ref{tab-apertures}.  The top and middle panel pairs to the
right show the average spectrum together with the dispersion, the full
range observed, and the median uncertainties derived from all
individual resolution elements for M\,82 and NGC\,253, respectively.
The resolution elements correspond to rebinned $2 \times 2$ pixels
for M\,82 and $4 \times 4$ pixels for NGC\,253
(i.e. $6^{\prime\prime} \times 6^{\prime\prime}$ in both cases).
The bottom right panel pair compares the spectra of the starburst 
core regions of each galaxy and their relative uncertainties.
The relative uncertainties are expressed as fraction of the 
measured flux density.
}
\label{fig-spec_stbs}
\end{figure*}

Figure~\ref{fig-selreg} indicates the selected regions on maps of the
$14.8 - 15.2~\mathrm{\mu m}$ emission.  Figure~\ref{fig-spec_stbs} shows
the spectra, all normalized to a total $6.0 - 6.6~\mathrm{\mu m}$ flux
density of unity to facilitate comparison of the relative strength
of the emission features at $\lambda \la 11~\mathrm{\mu m}$ and of
the continuum intensity at longer wavelengths.
The formal effective uncertainties are plotted along with the spectra.
Table~\ref{tab-apertures} gives the parameters for the synthetic apertures,
the normalizing fluxes, and the average and median uncertainties for each
spectrum.  Figure~\ref{fig-spec_stbs} also shows for \object{M\,82}
and \object{NGC\,253} the average
spectrum of all individual resolution elements within the ISOCAM field
of view, normalized as described above, together with the dispersion around
the mean, the full range observed, and the typical (median) uncertainties.
The resolution elements for \object{M\,82} and
\object{NGC\,253} correspond to rebinned pixels of size
$2 \times 2$ and $4 \times 4$ in original detector pixels, respectively.

All spectra look very similar and exhibit the ``classical''
characteristics observed towards star-forming regions and galaxies:
conspicuous broad emission features (the ``unidentified infrared bands''
or UIBs), a featureless continuum rising importantly at
$\lambda \ga 11~\mathrm{\mu m}$, and an apparent dip near 10$\,\mu$m.
The UIBs are attributed to a family of particles, the nature of which
still is debated, stochastically heated by single ultraviolet photons
while the long-wavelength continuum is ascribed to very small dust
grains between the transient heating and thermal equilibrium regimes
depending on grain properties and radiation field intensity
(e.g. L\'eger et al.~\cite{Leg89}; Allamandola et al.~\cite{All89};
D\'esert et al.~\cite{Des90}; Duley \& Williams \cite{Dul91};
Tielens et al.~\cite{Tie99};
see also reviews by Puget \& L\'eger \cite{Pug89};
Cesarsky \& Sauvage \cite{Ces99}; Genzel \& Cesarsky \cite{Gen00}).
We will hereafter refer to these ISM components as ``PAHs,''
adopting the currently popular model in which the UIB carriers consist
of polycyclic aromatic hydrocarbon molecules, and ``VSGs.''

\begin{figure*}[!ht]
\centering
\resizebox{0.85\hsize}{!}
          {\includegraphics[bb=17 105 578 625,clip]{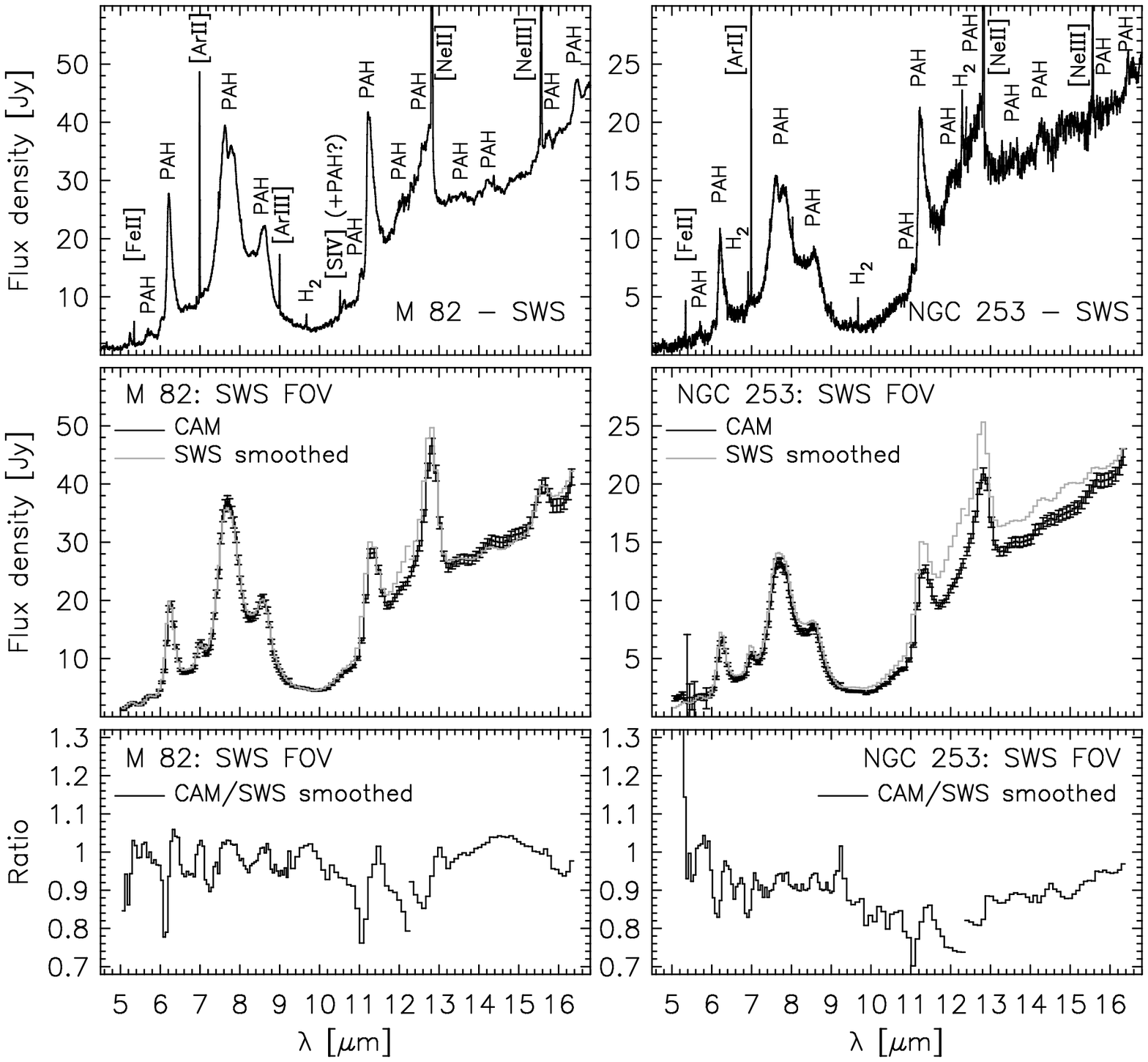}}
\caption
{
Comparison of MIR spectra of M\,82 and NGC\,253 obtained with ISOCAM
and SWS.  The top panels show the full resolution SWS spectra ($R \sim 1000$
to 500 from short to long wavelengths), with the identification of the
detected emission features (from F\"orster Schreiber et al.~\cite{FS01} 
for M\,82, and Sturm et al.~\cite{Stu00} for NGC\,253).  
For M\,82, the peak flux densities for the [NeII] $12.81~\mathrm{\mu m}$
and [NeIII] $15.56~\mathrm{\mu m}$ lines are 291.8~Jy and 109.1~Jy.
For NGC\,253, those for the [ArII] $6.99~\mathrm{\mu m}$,
[NeII] $12.81~\mathrm{\mu m}$, and [NeIII] $15.56~\mathrm{\mu m}$ lines
are 31.2~Jy, 140.6~Jy, and 35.2~Jy.  
The middle panels compare the SWS data convolved to the resolution of
ISOCAM ($R \sim 35$ to 45 from short to long wavelengths) with the ISOCAM
spectra (with uncertainties) integrated over the SWS aperture, while the
bottom panels show the ratio of the ISOCAM and smoothed SWS spectra.
In the spectral range covered by ISOCAM, the SWS aperture changes
from $14^{\prime\prime} \times 20^{\prime\prime}$ 
shortwards of $\lambda \approx 12.0~\mathrm{\mu m}$
to $14^{\prime\prime} \times 27^{\prime\prime}$ at longer wavelengths.
The long wavelength portions of the spectra have been scaled to match the
smaller SWS apertures by factors of 0.83 for M\,82 and 0.9 for NGC\,253,
as determined from the continuum level measured in the ISOCAM data.
}
\label{fig-spec_sws}
\end{figure*}

A number of weaker emission features are also detected in the spectra of
Fig.~\ref{fig-spec_stbs}.  However, their identification is problematic
at such low spectral resolution due to possible feature blends.
To emphasize this point and secure the identifications, 
Fig.~\ref{fig-spec_sws} shows the spectra of \object{M\,82} and
\object{NGC\,253} obtained at $R \sim 500 - 1000$ with the {\em ISO\/}-SWS
(from Sturm et al.~\cite{Stu00} and F\"orster Schreiber et al.~\cite{FS01}).
The SWS data are also compared with ISOCAM spectra taken in the
same apertures, after convolution to the same spectral resolution.
The ISOCAM and SWS data agree very well, confirming the accuracy
of the absolute and relative flux calibration for both instruments.
For \object{M\,82}, the differences are 5\% on average (24\% at most).
For \object{NGC\,253}, they are of 11\% on average (30\% at most) at
$\lambda \geq 5.4~\mathrm{\mu m}$ while they reach a factor of two at
shorter wavelengths, probably due to residual transient memory effects.

From this comparison (see also Sturm et al.~\cite{Stu00}), the features
at 6.2, 7.7, 8.6, and 11.3$\,\mu$m are unambiguously identified with PAH
emission.  The broad feature near 12.7$\,\mu$m clearly results from the
blending of the PAH 12.7$\,\mu$m band and of the [\ion{Ne}{ii}] 12.81$\,\mu$m
fine-structure line.
The [\ion{Ne}{iii}] 15.56$\,\mu$m line is also blended with the nearby
PAH 15.7$\,\mu$m feature.  From the SWS data, and with the continuum and
integration bandpasses given below, the PAH 12.7$\,\mu$m accounts for
about 50\% of the flux in the 12.7$\,\mu$m blend for both galaxies.
The PAH 15.7$\,\mu$m contains about 30\% of the flux in the 15.6$\,\mu$m
blend for \object{M\,82} (the noisier SWS spectrum of \object{NGC\,253}
makes an estimate difficult).  The least contaminated fine-structure line
detected with ISOCAM is [\ion{Ar}{ii}] 6.99$\,\mu$m, with the underlying PAH
at 7.0$\,\mu$m and the $\mathrm{H_{2}}$ $0 - 0~S(5)$ rotational line at
6.91$\,\mu$m contributing
$\approx 20\%$ and 5\%, respectively, to the blend flux in both galaxies.
The weak features at 5.65, 13.55, and 14.25$\,\mu$m are identified with
PAH bands; the latter is definitely not due to the high excitation
[\ion{Ne}{v}] 14.32$\,\mu$m line which is undetected in the SWS data.
The final identifications in the ISOCAM spectra are given in 
Fig.~\ref{fig-spec_stbs}.  Because of the excellent correspondence between
features seen in all three galaxies, the identifications for \object{M\,82}
and \object{NGC\,253} are adopted for \object{NGC\,1808} as well.

\begin{table*}[!ht]
\caption[]{Synthetic apertures for selected regions  \label{tab-apertures}}
\setlength{\tabcolsep}{0.25cm}
\begin{center}
\begin{tabular}{llllllll}
\hline\hline
Source & Region & Position & 
$\Delta\alpha$ $^{a}$ & 
$\Delta\delta$ $^{a}$ & 
Aperture & $f_\mathrm{6.0-6.6\,\mu m}$ $^{b}$ & Uncertainties $^{c}$ \\
   &   &   & (arcsec) & (arcsec) & (arcsec) & (Jy) & (\%) \\
\hline
M\,82     & CAM FOV  & ... & \ \ 0.0 & $+5.0$ & 
  $81 \times 81$ at P.A. of $212^{\circ}$ & $45.0 \pm 0.3$ & 6 (1) \\
      & Disk  & Nucleus  & \ \ 0.0 & \ \ 0.0 & $20 \leq r \leq 30$ & 
        $9.37 \pm 0.06$ & 3 (2) \\
      & Core     & Nucleus  & \ \ 0.0 & \ \ 0.0 & $r \leq 15$ & 
        $19.5 \pm 0.2$ & 2 (2) \\
      & MIR peak & MIR Peak & $-3.0$ & $-1.5$ & $r \leq 6$ & 
        $5.74 \pm 0.10$ & 5 (5) \\
\hline
NGC\,253  & CAM FOV  & ... & $-4.0$ & $+1.0$ & 
  $40.5 \times 40.5$ at P.A. $290^{\circ}$ & $8.34 \pm 0.73$ & 20 (3) \\
      & Disk  & Nucleus  & \ \ 0.0 & \ \ 0.0 & $10 \leq r \leq 15$ & 
        $1.75 \pm 0.09$ & 11 (4) \\
      & Core     & Nucleus  & \ \ 0.0 & \ \ 0.0 & $r \leq 7.5$ & 
        $3.90 \pm 0.05$ & 4 (3) \\
      & MIR peak & MIR Peak & \ \ 0.0 & \ \ 0.0 & $r \leq 3$ & 
        $1.14 \pm 0.03$ & 8 (7) \\
\hline
NGC\,1808 & Core     & Nucleus  & \ \ 0.0 & \ \ 0.0 & $r \leq 15$ & 
        $1.77 \pm 0.15$ & 57 (4) \\
      & MIR peak & MIR Peak & $+1.5$ & $-1.5$ & $r \leq 6$ & 
        $0.74 \pm 0.02$ &  9 (8) \\
\hline
\end{tabular}
\begin{list}{}{}
\item[$^{a}$] Right ascension and declination offsets relative to
the position of the nucleus of each galaxy.
For \object{M\,82}: $\alpha_{2000} = 09^\mathrm{h} 55^\mathrm{m} 52\fs 2$,
$\delta_{2000} = +69^{\circ} 40^{\prime} 46\farcs 6$ (Dietz et al.~\cite{Die86}).
For \object{NGC\,253}: $\alpha_{2000} = 00^\mathrm{h} 47^\mathrm{m} 33\fs 1$,
$\delta_{2000} = -25^{\circ} 17^{\prime} 18\farcs 3$ (Sams et al.~\cite{Sam94}).
For \object{NGC\,1808}: $\alpha_{2000} = 05^\mathrm{h} 07^\mathrm{m} 42\fs 3$,
$\delta_{2000} = -37^{\circ} 30^{\prime} 46\farcs 3$
(Collison et al.~\cite{Col94}).
\item[$^{b}$] Integrated flux density between 6.0 and 6.6$\,\mu$m used
for the normalization of the spectra plotted in Fig.~\ref{fig-spec_stbs}.
The integration area reported for the ISOCAM field of view of \object{M\,82}
and \object{NGC\,253} excludes the edges which were not illuminated or are
highly noisy. 
Quoted uncertainties are computed from the formal effective uncertainties
for individual pixels and wavelength channels (see Sect.~\ref{Sect-obs}).
\item[$^{c}$] Average of the formal effective uncertainties per wavelength
channel for the integrated spectra (median values are given in parenthesis).
\end{list}
\end{center}
\end{table*}

We performed various continuum and feature measurements on the spectra.
Table~\ref{tab-defs} gives the intervals used in the computations and
tables~\ref{tab-contfluxes} and \ref{tab-featfluxes} report the results.
The flux densities in the $6.0 - 9.0~\mathrm{\mu m}$ and 
$13.5 - 15.0~\mathrm{\mu m}$
bands are dominated by emission from PAHs and VSGs, respectively.
We chose the narrow continuum bands centered at 5.5 and 15.0$\,\mu$m
so as to minimize the contribution from PAHs and other emission lines.
In particular, we obtain the same 5.5$\,\mu$m flux densities within
$\approx 5\%$ from the ISOCAM data and from the higher resolution SWS
spectra for the SWS field of view in both \object{M\,82} and
\object{NGC\,253}, indicating negligible contribution from the adjacent
PAH 5.65$\,\mu$m feature and [\ion{Fe}{ii}] 5.34$\,\mu$m line in the
lower resolution data.
We measured the fluxes in the PAH features at 6.2, 7.7, 8.6,
and 11.3$\,\mu$m, in the [\ion{Ar}{ii}] 6.99$\,\mu$m line, and in 
the PAH 12.7$\,\mu$m $+$ [\ion{Ne}{ii}] 12.81$\,\mu$m and
PAH 15.7$\,\mu$m $+$ [\ion{Ne}{iii}] 15.56$\,\mu$m blends by
integrating the flux under the feature profiles after subtracting 
a continuum baseline obtained by linear interpolation between adjacent
spectral intervals.  More sophisticated methods such as profile fitting 
(e.g. Uchida et al.~\cite{Uch00}) are not necessary for our
purposes and are difficult to apply to the lower signal-to-noise 
(S/N) ratio data of individual pixels in generating linemaps
(Sect.~\ref{Sub-images}).

We did not compute fluxes for the individual [\ion{Ne}{ii}] 12.81$\,\mu$m
and [\ion{Ne}{iii}] 15.56$\,\mu$m lines.  The wavelength sampling is
too coarse for reliable profile decomposition, with the feature peaks
of each blend sampled by adjacent wavelength channels.  In addition, our
attempts to subtract the PAH contribution by attributing an excess in the
blend\,12.7$\,\mu$m/PAH\,11.3$\,\mu$m ratio to the [\ion{Ne}{ii}]
12.81$\,\mu$m line proved too sensitive to the definition of ``pure''
PAH~12.7$\,\mu$m/11.3$\,\mu$m ratio (e.g. as measured outside
of the starburst cores where comparatively little fine-structure line
emission from \ion{H}{ii} regions is expected).  Complications further arise
from possible intrinsic variations in the PAH ratios, extinction effects
($A_\mathrm{12.7\,\mu m}/A_\mathrm{11.3\,\mu m} \approx 0.55$), and
unconstrained fine-structure line emission from disk \ion{H}{ii} regions.
The fluxes for [\ion{Ar}{ii}] 6.99$\,\mu$m are much more reliable because
possible contributions by other features in our sample galaxies are
substantially smaller, as mentioned above.

\begin{table*}[!ht]
\caption[]{Intervals used to measure the continuum and features 
            \label{tab-defs}}
\setlength{\tabcolsep}{0.60cm}
\begin{center}
\begin{tabular}{llll}
\hline\hline
Property & Symbol & \multicolumn{1}{c}{Continuum points or intervals} & 
\multicolumn{1}{c}{Integration limits} \\
         &  & \multicolumn{1}{c}{($\,\mu$m)} & \multicolumn{1}{c}{($\,\mu$m)} \\ 
\hline
\ PAH emission~$^{a}$  & $f_\mathrm{PAH}$ & ... & $6.00 - 9.00$ \\
\ VSG emission~$^{b}$  & $f_\mathrm{VSG}$ & ... & $13.5 - 15.0$ \\
\ 5.5$\,\mu$m continuum  & $f_\mathrm{5.5\,\mu m}$ & ... & $5.40 - 5.52$ \\
\ 15.0$\,\mu$m continuum  & $f_\mathrm{15\,\mu m}$ & ... & $14.8 - 15.2$ \\
\ PAH 6.2$\,\mu$m & $F_{6.2}$ & $5.81 - 5.99$, $6.57 - 6.80$ & $6.04 - 6.51$ \\
\ PAH 7.7$\,\mu$m & $F_{7.7}$ & 7.14, $8.22 - 8.38$ & $7.19 - 8.17$ \\
\ PAH 8.6$\,\mu$m & $F_{8.6}$ & $8.27 - 8.33$, $8.84 - 8.89$ & $8.33 - 8.84$ \\
\ PAH 11.3$\,\mu$m & $F_{11.3}$ & 10.95, $11.7 - 11.8$ & $11.1 - 11.7$ \\
\ PAH 12.7$\,\mu$m + [\ion{Ne}{ii}] 12.81$\,\mu$m & $F_{12.7}$ & 
$12.1 - 12.2$, $13.2 - 13.3$ & $12.3 - 13.2$ \\
\ PAH 15.7$\,\mu$m + [\ion{Ne}{iii}] 15.56$\,\mu$m & $F_{15.6}$ & 
$15.0 - 15.2$, $16.0 - 16.1$ & $15.3 - 15.9$ \\
\ [\ion{Ar}{ii}] 6.99$\,\mu$m & $F_\mathrm{[Ar\,II]}$ & 
$6.74 - 6.86$, $7.14 - 7.19$ & $6.86 - 7.14$ \\
\hline
\end{tabular}
\begin{list}{}{}
\item[$^{a}$] Integrated flux in the band including the 
short wavelength PAH emission complex.
\item[$^{b}$] Integrated flux in the band probing the continuum
emission from VSGs free from strong emission lines and PAH features.
\end{list}
\end{center}
\end{table*}

\begin{table*}
\caption[]{Broad- and narrow-band measurements~$^{a}$
           \label{tab-contfluxes}}
\setlength{\tabcolsep}{0.30cm}
\begin{center}
\begin{tabular}{llllll}
\hline\hline
Source & Region & $f_\mathrm{PAH}$ & $f_\mathrm{VSG}$ &
 $f_\mathrm{5.5\,\mu m}$ & $f_\mathrm{15\,\mu m}$ \\
\hline
M\,82 
 & CAM FOV & $61.1 \pm 0.8$ & $78.3 \pm 0.3$ & $6.19 \pm 1.98$ & $80.8 \pm 0.6$ \\
 & Disk & $12.7 \pm 0.1$ & $11.3 \pm 0.1$ & $1.16 \pm 0.17$ & $10.8 \pm 0.1$ \\
 & Core & $26.9 \pm 0.1$ & $45.7 \pm 0.3$ & $3.15 \pm 0.08$ & $49.3 \pm 0.5$ \\
 & MIR Peak & $8.03 \pm 0.06$ & $16.3 \pm 0.2$ & $1.02 \pm 0.04$ & $17.9 \pm 0.4$ \\
\hline
NGC\,253 
 & CAM FOV & $11.9 \pm 0.3$ & $25.3 \pm 0.2$ & $2.12 \pm 1.31$ & $27.2 \pm 0.3$ \\
 & Disk & $2.49 \pm 0.03$ & $3.93 \pm 0.03$ & $0.41 \pm 0.19$ & $4.12 \pm 0.06$ \\
 & Core & $5.63 \pm 0.03$ & $15.3 \pm 0.1$ & $1.05 \pm 0.11$ & $16.8 \pm 0.3$ \\
 & MIR Peak & $1.68 \pm 0.02$ & $5.74 \pm 0.09$ & $0.35 \pm 0.02$ & $6.40 \pm 0.19$ \\
\hline
NGC\,1808 
 & Core & $2.66 \pm 0.05$ & $3.13 \pm 0.04$ & ... & $3.08 \pm 0.07$ \\
 & MIR Peak & $1.10 \pm 0.01$ & $1.39 \pm 0.03$ & $0.11 \pm 0.01$ & $1.41 \pm 0.05$ \\
\hline
\end{tabular}
\begin{list}{}{}
\item[$^{a}$] All flux densities are expressed in Jy.
The uncertainties result from the formal effective
uncertainties of the relative fluxes (see Sect.~\ref{Sect-obs}).
\end{list}
\end{center}
\end{table*}

\begin{table*}
\caption[]{Emission feature measurements~$^{a}$
           \label{tab-featfluxes}}
\setlength{\tabcolsep}{0.15cm}
\begin{center}
\begin{tabular}{lllllllll}
\hline\hline
Source & Region & $F_{6.2}$ & $F_{7.7}$ & $F_{8.6}$ & $F_{11.3}$
   & $F_{12.7}$ & $F_{15.6}$ & $F_\mathrm{[Ar\,II]}$ \\
\hline
M\,82
 & CAM FOV & $103 \pm 1$ & $226 \pm 1$ & $31.8 \pm 2.4$ & $57.7 \pm 0.9$
  & $54.0 \pm 0.4$ & $5.77 \pm 0.47$ & $6.05 \pm 0.31$ \\
 & Disk & $21.5 \pm 0.2$ & $46.5 \pm 0.2$ & $6.70 \pm 0.08$ & $13.1 \pm 0.1$
  & $9.74 \pm 0.06$ & $0.76 \pm 0.02$ & $0.67 \pm 0.02$ \\
 & Core & $43.7 \pm 0.5$ & $100 \pm 1$ & $13.8 \pm 0.4$ & $22.0 \pm 0.5$
  & $26.4 \pm 0.5$ & $3.49 \pm 0.39$ & $3.96 \pm 0.15$ \\
 & MIR Peak & $12.8 \pm 0.3$ & $29.7 \pm 0.4$ & $3.94 \pm 0.15$ & $5.76 \pm 0.25$
  & $8.50 \pm 0.35$ & $1.43 \pm 0.21$ & $1.64 \pm 0.08$ \\
\hline
NGC\,253
 & CAM FOV & $16.1 \pm 1.1$ & $36.3 \pm 0.3$ & $4.24 \pm 0.15$ & $10.2 \pm 0.2$
  & $10.6 \pm 0.2$ & $0.55 \pm 0.12$ & $1.93 \pm 0.12$ \\
 & Disk & $3.55 \pm 0.09$ & $7.63 \pm 0.07$ & $1.03 \pm 0.02$ & $2.41 \pm 0.03$
  & $2.02 \pm 0.02$ & $0.077 \pm 0.007$ & $0.36 \pm 0.01$ \\
 & Core & $7.30 \pm 0.11$ & $17.6 \pm 0.2$ & $1.86 \pm 0.05$ & $4.09 \pm 0.11$
  & $5.71 \pm 0.17$ & $0.37 \pm 0.10$ & $1.13 \pm 0.03$ \\
 & MIR Peak & $2.02 \pm 0.06$ & $4.95 \pm 0.09$ & $0.54 \pm 0.02$ & $1.07 \pm 0.06$
  & $1.83 \pm 0.10$ & $0.19 \pm 0.05$ & $0.37 \pm 0.02$ \\
\hline
NGC\,1808
  & Core & $4.17 \pm 0.19$ & $9.39 \pm 0.12$ & $1.38 \pm 0.03$ & $3.32 \pm 0.08$
   & $3.05 \pm 0.06$ & $0.096 \pm 0.016$ & $0.23 \pm 0.02$ \\
  & MIR Peak & $1.59 \pm 0.05$ & $3.74 \pm 0.08$ & $0.57 \pm 0.02$ & $1.31 \pm 0.05$
   & $1.34 \pm 0.04$ & $0.039 \pm 0.008$ & $0.087 \pm 0.007$ \\
\hline
\end{tabular}
\begin{list}{}{}
\item[$^{a}$] All fluxes are expressed in $10^{-14}~\mathrm{W\,m^{-2}}$.
The uncertainties result from the formal effective uncertainties
of the relative fluxes (see Sect.~\ref{Sect-obs}).
\end{list}
\end{center}
\end{table*}

\subsection{Mid-infrared images: mapping the features}  \label{Sub-images}

\begin{figure*}[!p]
\centering
\resizebox{0.85\hsize}{!}
          {\includegraphics[bb=17 45 578 780,clip]{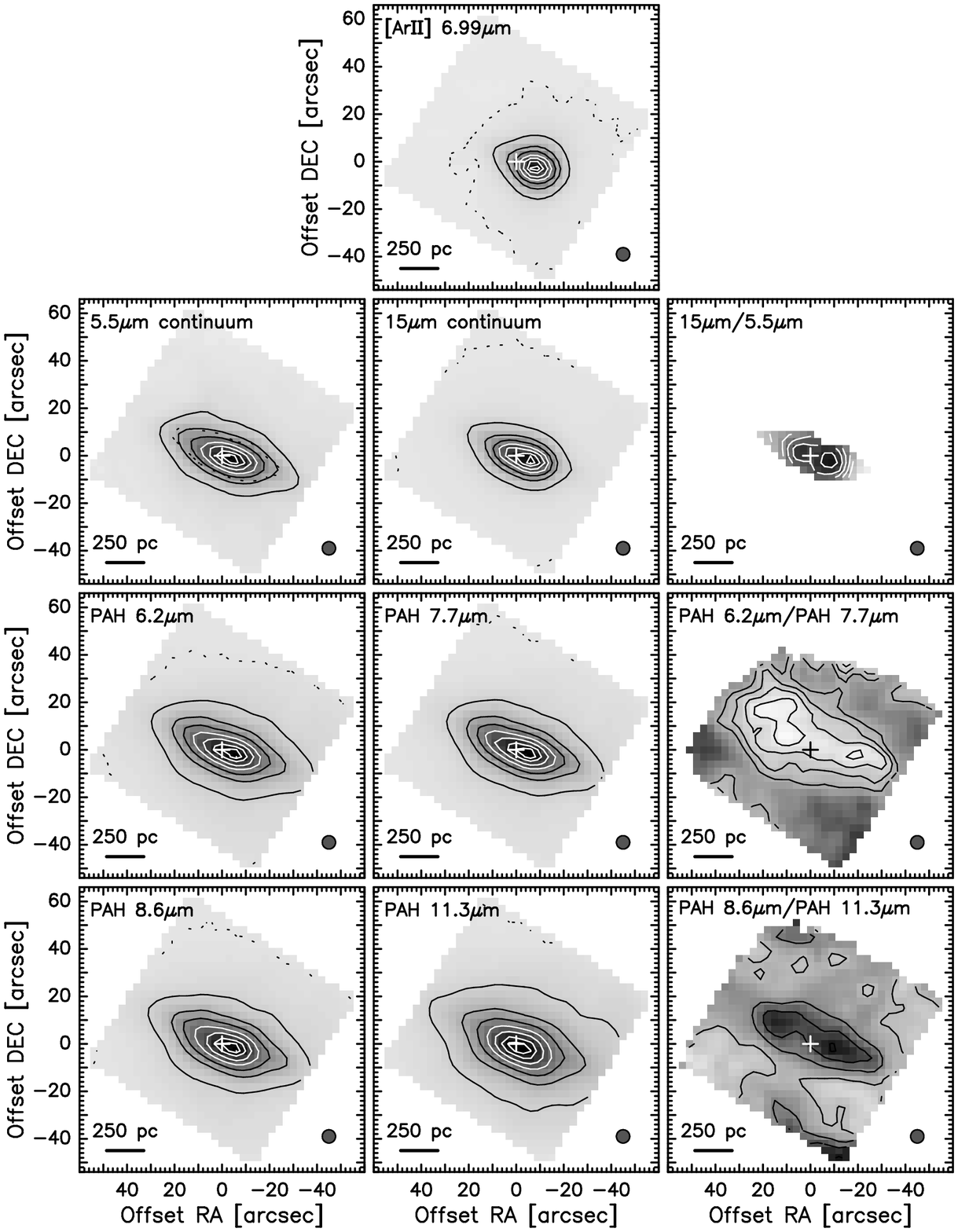}}
\caption
{
Selected ISOCAM maps of M\,82.
The maps are identified at the top of each panel while the physical linear
scale and the FWHM of the PSF are indicated at the bottom left and right.
The grayscale is linear, with white and black tones corresponding
to the minimum and maximum values displayed.
For the continuum and emission feature images, the solid contours 
are 10\%, 25\%, 40\%, 55\%, 70\%, 85\%, and 95\% of the maximum fluxes.  
For the 5.5 and $15.0~\mathrm{\mu m}$ continuum maps, the peaks are
10.3 and $187~\mathrm{mJy\,arcsec^{-2}}$, respectively.
For the [ArII] $6.99~\mathrm{\mu m}$ and the PAH 6.2, 7.7, 8.6,
and 11.3$\,\mu$m maps, they are 2.05, 12.5, 28.9, 3.84, and 5.54,
respectively, in units of $10^{-16}~\mathrm{W\,m^{-2}\,arcsec^{-2}}$.
The contour levels for the ratio maps are as follows:
from 10 to 18 in steps of 2 for the $\mathrm{15.0\,\mu m/5.5\,\mu m}$ 
continuum ratio, from 0.42 to 0.48 in steps of 0.02 for the 
PAH~$\mathrm{6.2\,\mu m/7.7\,\mu m}$ ratio,
and from 0.45 to 0.75 in steps of 0.10 for the 
PAH~$\mathrm{8.6\,\mu m/11.3\,\mu m}$ ratio.
The dotted contours in the continuum and feature maps enclose the
regions where data values exceed $3\sigma$, and ratio maps are displayed
for the area where this is satisfied by both images involved.
The cross at relative coordinates $(0^{\prime\prime}, 0^{\prime\prime})$
indicates the location of the galaxy nucleus.
}
\label{fig-M82_maps}
\end{figure*}

\begin{figure*}[!ht]
\centering
\resizebox{0.85\hsize}{!}
          {\includegraphics[bb=17 45 578 600,clip]{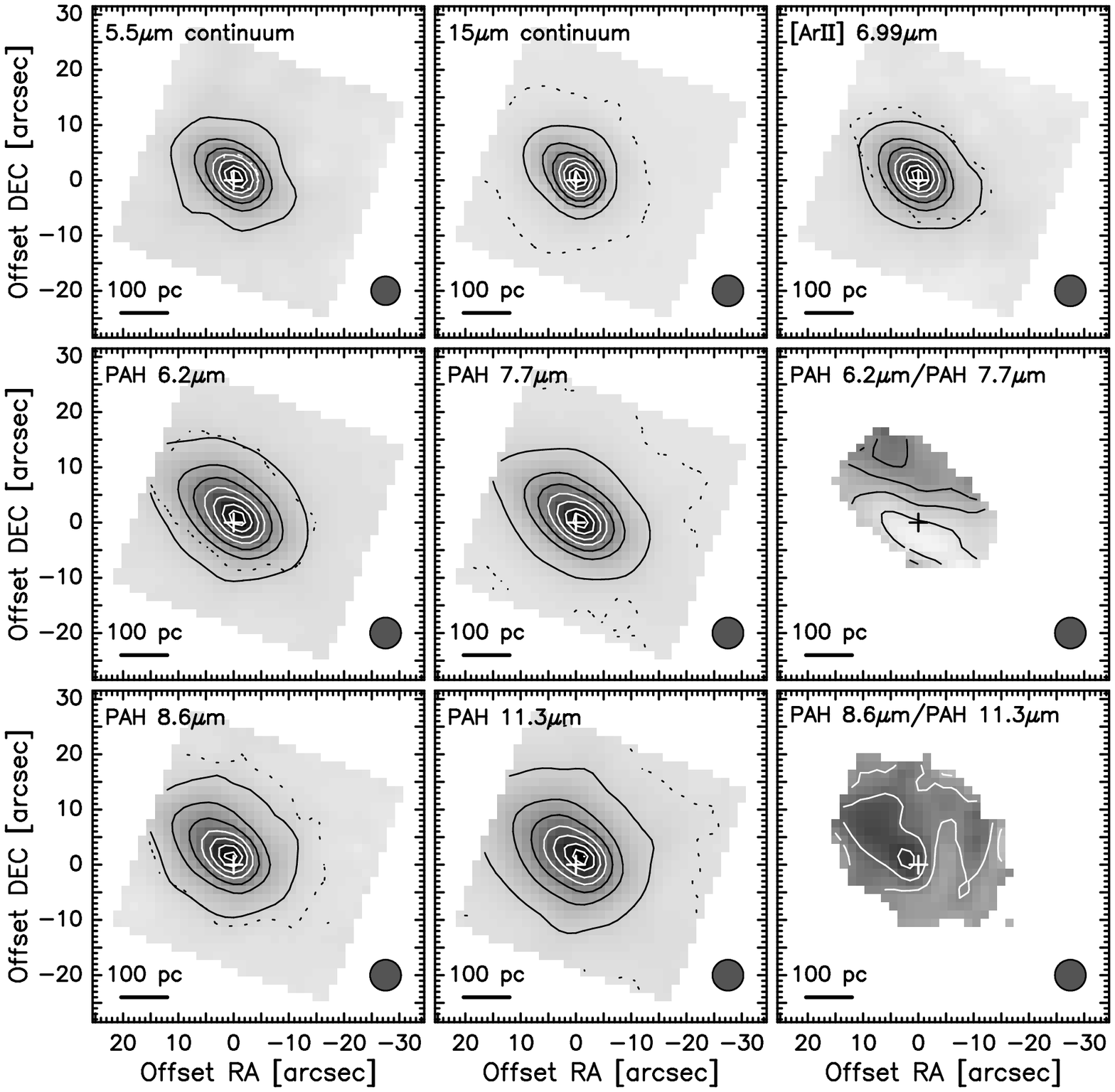}}
\caption
{
Selected ISOCAM maps of NGC\,253.
The maps are identified at the top of each panel while the physical linear
scale and the FWHM of the PSF are indicated at the bottom left and right.
The grayscale is linear, with white and black tones corresponding
to the minimum and maximum values displayed.
For the continuum and emission feature images, the contours are
10\%, 25\%, 40\%, 55\%, 70\%, 85\%, and 95\% of the maximum fluxes.  
For the 5.5 and $15.0~\mathrm{\mu m}$ continuum maps, 
the peaks are 13.6 and $269~\mathrm{mJy\,arcsec^{-2}}$.
For the [ArII] $6.99~\mathrm{\mu m}$ and the PAH 6.2, 7.7, 8.6, 
and $11.3~\mathrm{\mu m}$ maps, they are 1.50, 7.66, 19.1, 2.33, and 4.08,
respectively, in units of $10^{-16}~\mathrm{W\,m^{-2}\,arcsec^{-2}}$.
The contour levels for the ratio maps are as follows:
from 0.40 to 0.55 in steps of 0.05 for the 
PAH~$\mathrm{6.2\,\mu m/7.7\,\mu m}$ ratio,
and from 0.30 to 0.60 in steps of 0.10 for the 
PAH~$\mathrm{8.6\,\mu m/11.3\,\mu m}$ ratio.
The dotted contours in the continuum and feature maps enclose the regions
where data values exceed $3\sigma$ (for the $5.5~\mathrm{\mu m}$ continuum
map, the $3\sigma$ limit lies between the contours of 55\% and 70\% of
maximum flux), and ratio maps are displayed for the area where this is
satisfied by both images involved.
The cross at relative coordinates $(0^{\prime\prime}, 0^{\prime\prime})$
indicates the location of the galaxy nucleus.
}
\label{fig-N253_maps}
\end{figure*}

\begin{figure*}[!ht]
\centering
\resizebox{0.85\hsize}{!}
          {\includegraphics[bb=17 45 578 600,clip]{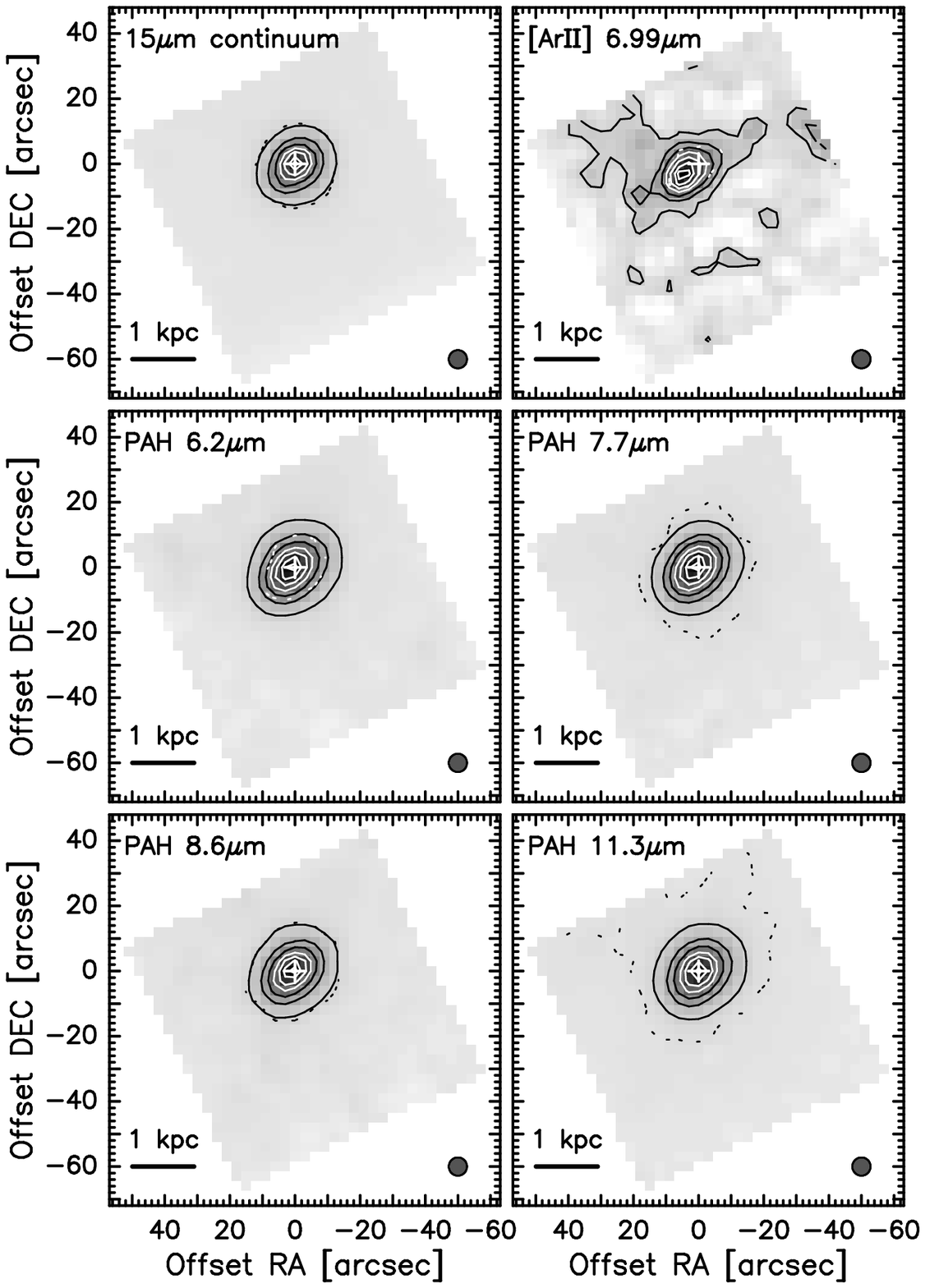}}
\caption
{
Selected ISOCAM maps of NGC\,1808.
The maps are identified at the top of each panel while the physical linear
scale and the FWHM of the PSF are indicated at the bottom left and right.
The grayscale is linear, with white and black tones corresponding
to the minimum and maximum values displayed.
The contours correspond to
10\%, 25\%, 40\%, 55\%, 70\%, 85\%, and 95\% of the maximum fluxes.
For the $15.0~\mathrm{\mu m}$ continuum map, the peak is 
$18.7~\mathrm{mJy\,arcsec^{-2}}$.
For the [ArII] $6.99~\mathrm{\mu m}$ and the PAH 6.2, 7.7, 8.6,
and 11.3$\,\mu$m maps, they are 0.13, 1.82, 4.23, 0.71, and 1.58,
respectively, in units of $10^{-16}~\mathrm{W\,m^{-2}\,arcsec^{-2}}$.
The dotted contours in the continuum and feature maps enclose the regions
where data values exceed $3\sigma$.
The cross at relative coordinates $(0^{\prime\prime}, 0^{\prime\prime})$
indicates the location of the galaxy nucleus.
}
\label{fig-N1808_maps}
\end{figure*}

We obtained broad- and narrow-band images as well as maps of the PAHs 
and [\ion{Ar}{ii}] 6.99$\,\mu$m line emission from the ISOCAM data cubes by
applying to each pixel the procedures described above for the spectra.  
Figures~\ref{fig-M82_maps}, \ref{fig-N253_maps}, and \ref{fig-N1808_maps}
present selected images and ratio maps for \object{M\,82},
\object{NGC\,253}, and \object{NGC\,1808}.
Contours corresponding to the same levels relative to the peak intensity
are plotted for all continuum and emission feature maps for ease of
comparison by visual inspection.  A $3\sigma$ contour is also shown
to delineate regions where observed small-scale structures are reliable.
In the following, we describe the various maps source by source;
their interpretation will be discussed in subsequent sections.

\subsubsection{M\,82}  \label{Sub-M82images}

The continuum and emission feature maps in \object{M\,82} show a globally
smooth spatial distribution, centered and peaking roughly 5\arcsec\ southwest
of the nucleus.  The PAH and 5.5$\,\mu$m continuum emission are the most
extended and symmetric, with disk-like isophotes elongated along the galactic
plane ($\mathrm{P.A.} \approx 70^{\circ}$).  In contrast, the 15$\,\mu$m 
continuum and [\ion{Ar}{ii}] 6.99$\,\mu$m line emission have more compact
distributions which are more asymmetric relative to the nucleus.
The 15$\,\mu$m/5.5$\,\mu$m ratio map outlines well the difference in
peak morphology between the short and long wavelength continuum.
The [\ion{Ar}{ii}] 6.99$\,\mu$m distribution is the most compact, with
centroid (determined from the emission out to a radius of 25\arcsec)
displaced 3.5\arcsec\ southwest of that of the PAH and continuum
emission, and showing only a slight extension towards the east.

Our ISOCAM maps provide an important complementary dataset to existing
MIR images in the literature, which were mostly obtained in different bands
or over limited regions (although with higher angular resolution up to
$\approx 1^{\prime\prime}$).  The distributions observed in our PAH maps
and for the PAH 3.29$\,\mu$m feature by Normand et al.~(\cite{Nor95}) and
Satyapal et al.~(\cite{Sat95}) at $\approx 1^{\prime\prime}$ resolution
are consistent with each other.   Our 15$\,\mu$m continuum map and the
19.2$\,\mu$m image of Telesco et al.~(\cite{Tel91}) are similar.
Maps of the $N$-band (10.8$\,\mu$m) and $11.8 - 13.0~\mathrm{\mu m}$
emission generated from the ISOCAM data cubes agree well with those
of Telesco et al.~(\cite{Tel91}) and Telesco \& Gezari (\cite{Tel92})
within the regions covered by the latter two images.

Our [\ion{Ar}{ii}] 6.99$\,\mu$m linemap globally resembles those of
other tracers of ionized gas from \ion{H}{ii} regions at mid- and
near-infrared wavelengths such as [\ion{Ne}{ii}] 12.81$\,\mu$m,
[\ion{Ar}{iii}] 8.99$\,\mu$m, [\ion{S}{iv}] 10.51$\,\mu$m, Br$\alpha$,
Br$\gamma$, and Pa$\beta$
(Achtermann \& Lacy \cite{Ach95}; Satyapal et al.~\cite{Sat95}).
At a resolution of $1^{\prime\prime} - 2^{\prime\prime}$, these
maps reveal a rich sub-structure dominated by prominent sources
$\approx 6^{\prime\prime}$ and 12\arcsec\ southwest of the nucleus
and $\approx 6^{\prime\prime}$ to the northeast (labeled W1, W2, and
E1 by Achtermann \& Lacy \cite{Ach95}).  While their intensity ratio
depends somewhat on the emission line considered, W1 and W2 are
together about three times brighter than E1.  Radial velocity
data of the [\ion{Ne}{ii}] 12.81$\,\mu$m and Br$\gamma$ emission
(Larkin et al.~\cite{Lar94}; Achtermann \& Lacy \cite{Ach95})
suggest that most sources reside in a nearly edge-on rotating
ring at radius coinciding in projection with W1 and E1 as well as 
along the stellar bar at larger radii, where the most recent
starburst episode took place about 5~Myr ago
(e.g. F\"orster Schreiber et al.~\cite{FS02}).
Within the positional uncertainties and resolution limitations,
the spatial distribution of our [\ion{Ar}{ii}] 6.99$\,\mu$m map
peaks between W1 and W2 and encompasses E1, and thus traces
well the youngest starburst regions.

Little differences are seen between the PAH maps but the
PAH~6.2$\,\mu$m/7.7$\,\mu$m and PAH~8.6$\,\mu$m/11.3$\,\mu$m
ratio maps reveal spatial variations at the
20\% and 60\% level, respectively.  The variations are significant
within the brighter emission regions along the disk, where the relative
uncertainties are $< 15\%$.  Structures seen towards the map edges are much
less reliable as they become comparable in amplitude to the uncertainties.
The PAH~6.2$\,\mu$m/7.7$\,\mu$m ratio is lower along the disk and reaches
minima on each side of the nucleus.  The overall morphology appears to curve
northwards away from the nucleus and extensions are hinted at above and below
the galactic plane.  Similar shape and spurs are observed notably in the
large-scale distribution of the molecular and ionized gas line emission
and of the radio continuum emission (e.g. Shen \& Lo \cite{She95};
Achtermann \& Lacy \cite{Ach95}; Wills et al.~\cite{Wil99}). 
The spatial variations in the PAH~8.6$\,\mu$m/11.3$\,\mu$m ratio match
roughly those of the PAH~6.2$\,\mu$m/7.7$\,\mu$m ratio, with higher
values along the disk and maxima flanking the nucleus.  Noticeably,
the western peak lies closer to the nucleus than the western 
PAH~6.2$\,\mu$m/7.7$\,\mu$m minimum, and the apparent curving and
large-scale extensions have no counterpart in the
PAH~8.6$\,\mu$m/11.3$\,\mu$m map.

\begin{figure}[!ht]
\centering
\resizebox{\hsize}{!}
          {\includegraphics[bb=17 215 340 605,clip]{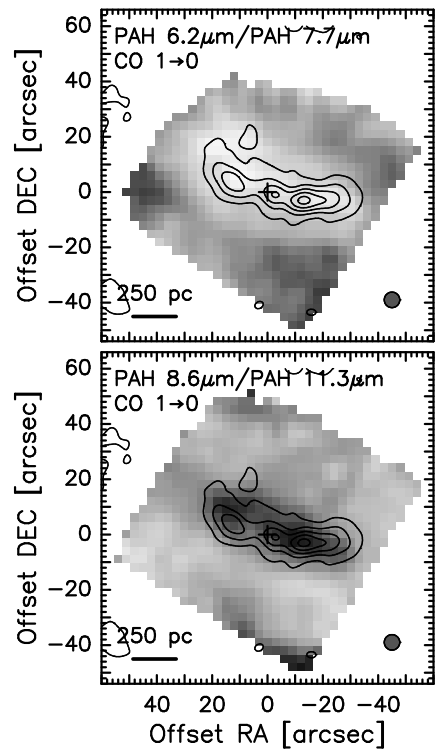}}
\caption
{
Comparison of ISOCAM PAH ratio maps of M\,82 with the distribution
of CO~$J : 1 \rightarrow 0$ millimetric emission.
The PAH~$\mathrm{6.2\,\mu m/7.7\,\mu m}$ and
PAH~$\mathrm{8.6\,\mu m/11.3\,\mu m}$ maps (top and bottom
panels, respectively) are displayed as grayscale images as for
Fig.~\ref{fig-M82_maps}.  
The contours show the CO~$J : 1 \rightarrow 0$ emission from the map of
Shen \& Lo (\cite{She95}) convolved at the resolution of the ISOCAM images
(from the original $2.5^{\prime\prime}$ to $5.6^{\prime\prime}$).
The physical linear scale and the FWHM of the ISOCAM PSF are indicated at the
bottom left and right.
The cross at relative coordinates $(0^{\prime\prime}, 0^{\prime\prime})$
marks the location of the galaxy nucleus.
}
\label{fig-M82_CAM_CO}
\end{figure}

Figure~\ref{fig-M82_CAM_CO} compares the PAH ratio maps with the
CO $J : 1 \rightarrow 0$ millimetric emission of Shen \& Lo (\cite{She95})
convolved at the resolution of the ISOCAM maps.  The overall correspondence
of the PAH~6.2$\,\mu$m/7.7$\,\mu$m minima and
PAH~8.6$\,\mu$m/11.3$\,\mu$m maxima
with the peaks in CO emission, as well as the curved shape and northeastern
extension for the PAH~6.2$\,\mu$m/7.7$\,\mu$m ratio, is quite striking.
We believe that the observed variations in PAH ratios are mostly real.
Characteristic patterns expected for ghosts are not seen (unresolved ring-
or arc-like features most prominent in the presence of point-like sources).
Artifacts due to the uncorrected flat field and straylight could produce
extended lobes on each side of an axis at
$\mathrm{P.A.} \approx 150^{\circ}$ for the \object{M\,82} data,
i.e. roughly the minor axis
(Biviano et al.~\cite{Biv98a, Biv98b}; Okumura \cite{Oku00}).
However, such lobes would have a much larger
extent than the features seen in our PAH ratio maps and the chromatic
dependence between 6.2 and 7.7$\,\mu$m, and 8.6 and 11.3$\,\mu$m is
predicted and observed to be smaller than the measured variations at
20\% and 60\% levels, respectively.

\subsubsection{NGC\,253}  \label{Sub-N253images}

In \object{NGC\,253}, the emission in the continuum bands, PAH features,
and [\ion{Ar}{ii}] 6.99$\,\mu$m line is characterized by a strong peak
within $1^{\prime\prime} - 2^{\prime\prime}$ of the nucleus,
embedded in a diffuse envelope elongated along the galactic plane 
($\mathrm{P.A.} \approx 50^{\circ}$).  The 15$\,\mu$m continuum and
[\ion{Ar}{ii}] 6.99$\,\mu$m line distributions do not seem to extend
as far as the PAH emission in the outer parts of the source.
The noisy $\lambda < 6~\mathrm{\mu m}$ channels prevent
reliable assessment of the lower level, large-scale 5.5$\,\mu$m
continuum emission.  The centroids in the various images are
essentially indistinguishable (differences $< 0.5~\mathrm{pixel}$).

Though the images reveal little spatial structure because of the
intrinsically small source size and the limited angular resolution,
they are consistent with expectations at this resolution based on
previously published MIR maps (mostly obtained at $\sim 1^{\prime\prime}$
resolution).  These include broad-band images at 10.8 ($N$ band) and
19.2$\,\mu$m, narrow-band images at 8.5, 10.0, 12.5, and 20.2$\,\mu$m
tracing PAH and/or continuum emission, maps of the 3.29 and 11.3$\,\mu$m
PAH emission, and images of the [\ion{Ne}{ii}] 12.81$\,\mu$m line and of
the underlying continuum $+$ PAH emission (Pi\~na et al.~\cite{Pin92}; 
Telesco et al.~\cite{Tel93}; Keto et al.~\cite{Ket93, Ket99};
Kalas \& Wynn-Williams \cite{Kal94}; B\"oker et al.~\cite{Bok98}).
We note that the [\ion{Ne}{ii}] linemaps of B\"oker et al.~(\cite{Bok98})
and Keto et al.~(\cite{Ket99}) likely contain a contribution from
PAH 12.7$\,\mu$m emission because the observations were made at low
spectral resolution ($\mathrm{FWHM \approx 0.2~\mu m}$).

The spatial distribution seen in all our maps is evidently dominated by a
very prominent compact MIR source, better outlined in arcsecond resolution
images.  Keto et al.~(\cite{Ket99}) associated this source with a bright
off-nucleus super star cluster resolved by optical
{\em Hubble Space Telescope\/} observations.  Based on their data, the
source has a size of $20~\mathrm{pc}$ (marginally resolved) and accounts for
$\approx 20\%$ of the total continuum emission at 12 and 20$\,\mu$m, and 12\%
of the total [\ion{Ne}{ii}] 12.81$\,\mu$m flux.  From our own
measurements, nearly 25\% of the total continuum emission detected
with ISOCAM between 13.5 and 15$\,\mu$m originates in the
70\,pc-diameter MIR peak; this fraction varies between 10\% and 30\%
for our various broad- and narrow-band and emission feature measurements
(tables~\ref{tab-contfluxes} and \ref{tab-featfluxes}).
\object{NGC\,253} is quite remarkable in the compactness of its main MIR
emitting region, with a $\mathrm{FWHM \la 150~pc}$ while the optical disk
extends over $\sim 20~\mathrm{kpc}$.  Although of a different nature, the
interacting system \object{NGC\,4038/4039} offers a similar example,
with 15\% of its $12.5 - 18~\mathrm{\mu m}$ luminosity being produced
within a 100\,pc-size star-forming knot in the overlapping region between
the galaxies (Mirabel et al.~\cite{Mir98}).

Additional small-scale structure of \object{NGC\,253}
at MIR wavelengths includes a secondary much fainter peak in the
12 and 20$\,\mu$m continuum emission nearly coinciding with the nucleus,
about 2\arcsec\ northeast of the prominent source discussed above.
The [\ion{Ne}{ii}] 12.81$\,\mu$m line emission differs somewhat from the
continuum, except possibly for the brightest peak, showing a bilobal or
arc-like structure also hinted at in Br$\gamma$ images and suggestive
of a circumnuclear star-forming ring (B\"oker et al.~\cite{Bok98};
Engelbracht et al.~\cite{Eng98}; Keto et al.~\cite{Ket99}).  None of
these features, however, is resolved with ISOCAM.

The PAH~6.2$\,\mu$m/7.7$\,\mu$m and PAH~8.6$\,\mu$m/11.3$\,\mu$m ratio
maps show variations of about 40\% and nearly a factor of 2, respectively.
These are statistically significant in view of the corresponding
formal uncertainties of $\leq 20\%$ and $< 30\%$.  The ratio maps
differ markedly, with the PAH~6.2$\,\mu$m/7.7$\,\mu$m image indicating
a general increase from south to north of the nucleus whereas the
PAH~8.6$\,\mu$m/11.3$\,\mu$m ratio appears more centrally concentrated,
peaking near the nucleus and slightly more extended northeast.
The reality of the arc-like feature $\approx 10^{\prime\prime}$
northwest of the nucleus in the PAH~8.6$\,\mu$m/11.3$\,\mu$m map is
dubious because it is barely resolved and could be due to ghost
effects given the strong unresolved peak of the emission.
No corresponding structure that could perhaps support this feature
is seen in maps of the molecular gas emission, of the radio continuum
emission, of tracers of ionized gas or even of extinction
(e.g. Peng et al.~\cite{Pen96};
Ulvestad \& Antonucci \cite{Ulv97}; Engelbracht et al.~\cite{Eng98}).
The limited region for which $\mathrm{S/N} \geq 3$ prevents us from
examining the larger-scale distribution in PAH ratios.

\subsubsection{NGC\,1808}  \label{Sub-N1808images}

The starburst in \object{NGC\,1808} covers a region of comparable physical
size to that of \object{M\,82} but being three times more distant, less
structural details are resolved by ISOCAM.
The PAH emission appears the most extended and oriented parallel to the
major axis of the galaxy ($\mathrm{P.A.} \approx 140^{\circ}$), peaking at
the nucleus.  The 15$\,\mu$m emission follows closely the PAH emission.  
The [\ion{Ar}{ii}] 6.99$\,\mu$m line emission is the most distinct in that
the peak clearly is off-nucleus, about 5\arcsec\ to the southeast, and the
centroid of the emission region lies 2.5\arcsec\ southeast of that for the
15$\,\mu$m continuum and PAH emission.  The [\ion{Ar}{ii}] 6.99$\,\mu$m map
agrees well in peak position and extent with the global distribution of the
most intense star-forming regions, or hot spots, as traced by H recombination
lines and radio continuum emission (e.g. Saikia et al.~\cite{Sai90};
Krabbe et al.~\cite{Kra94}; Kotilainen et al.~\cite{Kot96}).
Due to the limited number of pixels with $\mathrm{S/N} \geq 3$,
no useful ratio maps could be made for \object{NGC\,1808}.

Previously published $N$ band images of \object{NGC\,1808}
(Telesco et al.~\cite{Tel93}; Krabbe et al.~\cite{Kra01})
show an overall similar morphology as our 15$\,\mu$m continuum
and PAH feature maps with, at $\approx 1^{\prime\prime}$ resolution,
a strong point-like source at the nucleus and a southeastern
extension covering the starburst regions.
ISOCAM broad-band LW4 (6$\,\mu$m) polarisation observations were presented
by Siebenmorgen et al.~(\cite{Sie01}) along with a CVF spectrum of
the central 25\arcsec\ of \object{NGC\,1808} which is essentially
identical in shape to ours of the starburst core (30\arcsec\ aperture;
Fig.~\ref{fig-spec_stbs}).  Siebenmorgen et al.~(\cite{Sie01}) assigned
all detected features to PAHs and successfully reproduced them with PAH
emission alone in their radiative transfer models.
As also noted by these authors, the contribution of nebular gas emission
lines blended with PAH features is uncertain for \object{NGC\,1808},
and we cannot constrain it using, e.g., SWS data as for \object{M\,82}
and \object{NGC\,253}.  However, our [\ion{Ar}{ii}] 6.99$\,\mu$m map
clearly differs from those of the PAH emission and is consistent with
the spatial distribution of the bulk of \ion{H}{ii} regions, 
supporting the idea that the 7$\,\mu$m feature is indeed dominated by
the [\ion{Ar}{ii}] line instead of PAHs.

\subsection{Salient features}          \label{Sub-salient}

We summarize here the most important aspects of the results presented above:
\begin{itemize}
\item{in terms of features present, the MIR spectral energy
      distribution is essentially identical among the three galaxies
      as well as within each of them,}
\item{the largest spectral variations are observed in the
      $\lambda \ga 11~\mathrm{\mu m}$ continuum intensity
      relative to the shorter wavelength emission, while the shape
      of the $\lambda = 5 - 11~\mathrm{\mu m}$ region is nearly invariant,}
\item{the PAH emission is spatially more extended than the 15$\,\mu$m
      continuum and [\ion{Ar}{ii}] 6.99$\,\mu$m line emission,}
\item{the [\ion{Ar}{ii}] 6.99$\,\mu$m line emission traces well the young
      starburst regions, and}
\item{the relative intensities of the PAH features can exhibit complex
      spatial variations.}
\end{itemize}

\section{Origin and spatial distribution of the PAH and continuum emission}  
          \label{Sect-Spat_distr}

The ubiquity of PAH features -- seen towards sources as diverse as
circumstellar regions, diffuse atomic clouds, \ion{H}{ii} regions,
molecular clouds, normal galaxies, starbursts, and ULIRGs -- and the
close association of the MIR VSG continuum emission with Galactic and
extragalactic actively star-forming sites now seem well established;
the {\em ISO\/} mission played a major role in this recognition
(see Geballe \cite{Geb97}; Tokunaga \cite{Tok97}; 
Tielens et al.~\cite{Tie99}; Cesarsky \& Sauvage \cite{Ces99};
Genzel \& Cesarsky \cite{Gen00} for reviews).
Spatial mapping of Galactic sources has revealed in more detail the origin
of the main MIR emission components in star-forming regions: PAH features
arise predominantly in the photodissociation regions (PDRs) at the interface
between \ion{H}{ii} regions and molecular clouds while the VSG 
$\lambda \ga 11~\mathrm{\mu m}$ continuum is more intense and steeper closer
to the exciting source (e.g. Verstraete et al.~\cite{Ver96};
Cesarsky et al.~\cite{Ces96a, Ces96b}; Cr\'et\'e et al.~\cite{Cre99}).
By extension, in star-forming galaxies, the emission components
are believed to trace PDRs and \ion{H}{ii} regions, respectively.
This is supported by spectral decomposition of starburst galaxies
(Tran \cite{Tra98}; Laurent et al.~\cite{Lau00}), and by photometric 
and spectrophotometric imaging of spiral galaxies and starburst systems
(e.g. Mirabel et al.~\cite{Mir98}; Roussel et al.~\cite{Rou01a}).
Low excitation fine-structure line emission such as
[\ion{Ar}{ii}] 6.99$\,\mu$m, [\ion{Ar}{iii}] 8.99$\,\mu$m,
[\ion{Ne}{ii}] 12.81$\,\mu$m, and [\ion{Ne}{iii}] 15.56$\,\mu$m
usually accompanies the long-wavelength VSG continuum and originates
primarily in \ion{H}{ii} regions (e.g. Sturm et al.~\cite{Stu00}).

\begin{figure*}[!ht]
\centering
\resizebox{0.85\hsize}{!}
          {\includegraphics[bb=17 90 578 685,clip]{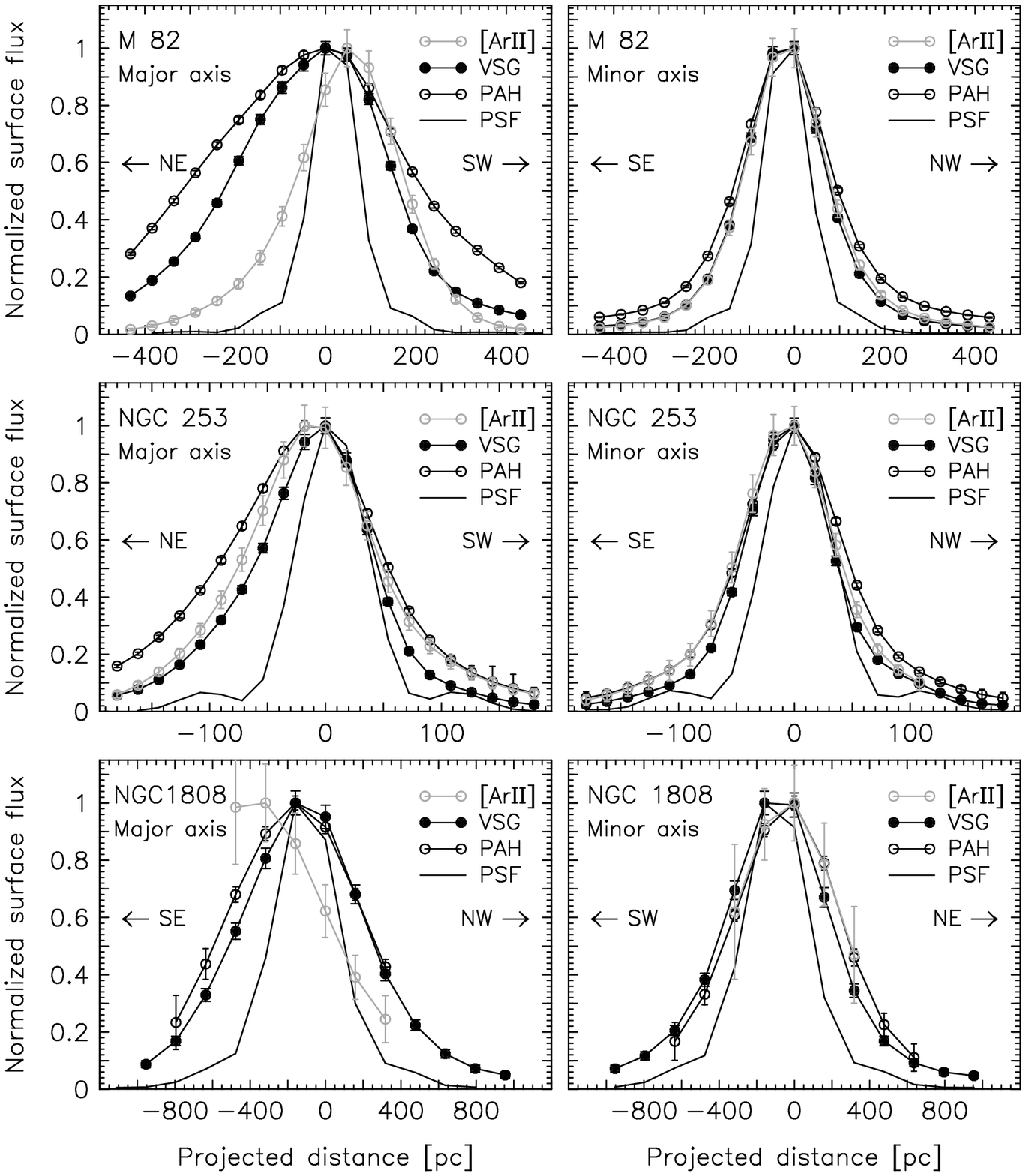}}
\caption
{
Major and minor axis light profiles for M\,82, NGC\,253, and NGC\,1808.
The curves represent the variations of the integrated flux per unit surface
area as a function of projected distance from the nucleus for the
[Ar\,\small{II}]~$6.99~\mathrm{\mu m}$ line
(grey lines and open circles), the $13.5 - 15.0~\mathrm{\mu m}$ VSG-dominated 
continuum (black lines and filled circles), and the $6.0 - 9.0~\mathrm{\mu m}$
PAH-dominated region (black lines and open circles).  The width of the
synthetic slits along each axis is 3 pixels for M\,82 and NGC\,1808
and 6 pixels for NGC\,253 (corresponding to 144, 477, and 108~pc,
respectively).  The curves are normalized to the maximum values
which are for M\,82, NGC\,253, and NGC\,1808, respectively:
0.26, 0.36, and 0.0036 $\mathrm{mJy\,pc^{-2}}$ for the 
``PAH'' major axis profiles,
0.28, 0.38, and 0.0035 $\mathrm{mJy\,pc^{-2}}$ for the 
``PAH'' minor axis profiles,
0.53, 1.20, and 0.0044 $\mathrm{mJy\,pc^{-2}}$ for the
``VSG'' major axis profiles,
0.57, 1.27, and 0.0044 $\mathrm{mJy\,pc^{-2}}$ for the
``VSG'' minor axis profiles,
6.32, 7.87, and 0.035 $10^{-19}~\mathrm{W\,m^{-2}\,pc^{-2}}$ for the
[Ar\,\small{II}] major axis profiles, and
5.77, 8.30, and 0.024 $10^{-19}~\mathrm{W\,m^{-2}\,pc^{-2}}$ for the
[Ar\,\small{II}] minor axis profiles.
The PSF profiles for the spatially-smoothed data cubes are shown 
as well (simple black lines; see Sect.~\ref{Sect-obs}).
}
\label{fig-profiles}
\end{figure*}

The interpretation of the ISOCAM maps of \object{M\,82}, 
\object{NGC\,253}, and \object{NGC\,1808} within the above framework is not
straightforward.  Figure~\ref{fig-profiles} complements the various maps with
light profiles of the $6 - 9~\mathrm{\mu m}$, $13.5 - 15~\mathrm{\mu m}$,
and [\ion{Ar}{ii}] 6.99$\,\mu$m emission taken along the major and minor
axes of the galaxies.  The profiles clearly support the extension along
the major axis indicated by the maps, and show that the emission in these
tracers along the minor axis is but marginally resolved and essentially
indistinguishable within each galaxy.

Somewhat surprisingly given their common origin in \ion{H}{ii} regions,
a close spatial correlation between the VSG continuum and the
[\ion{Ar}{ii}] 6.99$\,\mu$m line emission is not observed in our sample
galaxies.  The exception may be \object{NGC\,253} but this probably results
from the intrinsically small MIR source size and resolution limitations
(Sect.~\ref{Sub-N253images}).  Although the extension to the northeast
is real, the differences between the various profiles even along the
major axis are smaller than the PSF FWHM.  We conclude for
\object{NGC\,253} that on scales comparable to and larger than the
resolution of our ISOCAM maps (5.2\arcsec, or $\approx 60~\mathrm{pc}$),
no significant differences are observed in the spatial distribution of the
PAH, VSG, and [\ion{Ar}{ii}] 6.99$\,\mu$m emission within the central 400~pc.

For \object{M\,82}, the [\ion{Ar}{ii}] 6.99$\,\mu$m emission along
the galactic plane peaks at a different position and is more compact
than the VSG and PAH emission.
The different distributions of the VSG and [\ion{Ar}{ii}] 6.99$\,\mu$m 
emission could be interpreted in terms of differences in average energy
of the exciting photons: the [\ion{Ar}{ii}] 6.99$\,\mu$m line traces the
$\geq 16~\mathrm{eV}$ radiation field while the VSGs may be heated by UV
photons at lower energies as well
(D\'esert et al.~\cite{Des90}; Siebenmorgen \& Kr\"ugel \cite{Sie92};
Boulanger et al.~\cite{Bou94}; Dwek et al.~\cite{Dwe97};
Jones \& d'Hendecourt \cite{Jon00}).
The brighter PAH emission around the nucleus could be originating
predominantly in the PDRs associated with the \ion{H}{ii} regions
within the most active starburst sites.  Beyond a radius of 
$\approx 300~\mathrm{pc}$, the PAH profile is resolved out from
the VSG profile and more extended.  These differences could
be attributed to excitation of aromatic band carriers by yet softer
photons or to an increased filling factor for the PDRs compared to
the \ion{H}{ii} regions in less intense starburst regions, and
perhaps to a contribution from diffuse cirrus clouds as observed
in the Milky Way and in some spiral galaxies
(e.g. Ristorcelli et al.~\cite{Ris94}; Onaka et al.~\cite{Ona96};
Mattila et al.~\cite{Mat99}; Roussel et al.~\cite{Rou01a};
Li \& Draine \cite{Li02}).
Alternative possibilities could include destruction by photodissociation
or depletion of the ISM in PAHs and VSGs in the most intense starburst
regions, flattening the spatial distribution around the peak
(e.g. Boulanger et al.~\cite{Bou88};
Carral et al.~\cite{Car94}; Normand et al.~\cite{Nor95}).

In \object{NGC\,1808}, the bulk of the [\ion{Ar}{ii}] 6.99$\,\mu$m emission is
clearly displaced to the southeast compared to the emission in the PAH and
VSG bands defined above, which have virtually identical distributions.
As discussed in Sect.~\ref{Sub-N1808images}, the [\ion{Ar}{ii}] 6.99$\,\mu$m
distribution agrees well with the off-nucleus location of the main
star-forming sites.  The spectrum integrated over the starburst core and
MIR peak in \object{NGC\,1808} is also much flatter than observed towards
the corresponding regions in \object{M\,82} and \object{NGC\,253} and
resembles more the spectrum of their outer disks
(Fig.~\ref{fig-spec_stbs}).  The long-wavelength continuum in
\object{NGC\,1808} may in fact not be dominated by VSG emission
from pure \ion{H}{ii} regions but rather produced primarily in
PDRs by small particles akin to the carriers of the main UIB bands.
This is reminiscent of the situation in disks of spiral galaxies.
There, the 15$\,\mu$m emission correlates with the 7$\,\mu$m
PAH-dominated emission and lies in a distinct regime compared to more
active regions including circumnuclear regions in spiral (especially
barred) galaxies and starbursts, characterized by an excess in the 
15$\,\mu$m/7$\,\mu$m colour (e.g. Roussel et al.~\cite{Rou01a};
Dale et al.~\cite{Dal01}; F\"orster Schreiber, Roussel, \& Sauvage, in prep.).
From radiative transfer modeling of \object{NGC\,1808}'s
$\sim 3 - 200~\mathrm{\mu m}$ spectrum,
Siebenmorgen et al.~(\cite{Sie01}) concluded that the MIR range is
dominated by PAHs and large dust grains (radii between 100 and 2400~\AA)
with negligible contribution by very small graphites and silicates (radii
of $5 - 80$~\AA).  The key feature of their models is the inclusion of
hot spots where large grains are heated locally by massive stars to higher
temperatures giving rise to the 25$\,\mu$m emission and gradually
contributing less towards shorter wavelengths where PAHs take over
as main emitters.

We note that the morphology of the [\ion{Ar}{ii}] 6.99$\,\mu$m emission
for \object{M\,82} and \object{NGC\,1808}, especially the off-nucleus peak 
position, reflects the spatial distribution of the \ion{H}{ii} regions and
not variations in the excitation state of the ionized gas whereby, e.g.,
[\ion{Ar}{iii}] emission would become important at the expense of
[\ion{Ar}{ii}].  As discussed in Sect.~\ref{Sub-M82images}, the \ion{H}{ii}
regions in \object{M\,82} are concentrated in a circumnuclear ring-like
structure at radius of $\approx 85~\mathrm{pc}$.  Direct tracers
such as near-infrared to radio H recombination lines, MIR Ne, Ar, and S
fine-structure lines, the \ion{He}{i} 2.06$\,\mu$m recombination line, and
radio thermal continuum all show a similar distribution characterized by peaks
flanking the nucleus, with the southwestern complex being the most intense,
and little emission at the nucleus (e.g. Satyapal et al.~\cite{Sat95};
Achtermann \& Lacy \cite{Ach95}; Seaquist et al.~\cite{Sea96};
F\"orster Schreiber et al.~\cite{FS01}).  In particular, the
[\ion{Ar}{iii}] 8.99$\,\mu$m map of Achtermann \& Lacy (\cite{Ach95})
tracing the $\geq 28~\mathrm{eV}$ radiation field clearly shows this bilobal
distribution and relative lack of emission at the nucleus.  Less information
is available for \object{NGC\,1808} but Br$\gamma$ imaging clearly reveals 
the \ion{H}{ii} region complexes to lie predominantly to the east and
southeast of the nucleus, thus indicating a circumnuclear location for
the starburst (Krabbe et al.~\cite{Kra94}; Kotilainen et al.~\cite{Kot96}).

Extinction effects could partly account for differences in the PAH, VSG, and
[\ion{Ar}{ii}] 6.99$\,\mu$m emission on small spatial scales (comparable to
the resolution elements in our ISOCAM data) but are not likely to affect the
relative concentrations.  The highest levels of obscuration are measured in
the near surroundings of the MIR peaks in all three galaxies and generally
decrease at larger distances (e.g. Larkin et al.~\cite{Lar94};
Engelbracht et al.~\cite{Eng98}; Krabbe et al.~\cite{Kra94}).  Depending
on the extinction law applicable in the $3 - 10~\mathrm{\mu m}$ region
(Sect.~\ref{Sub-ext_effects} below), such a radial gradient in extinction
would either affect very little the relative global distributions of the
tracers discussed here, or suppress more severely the shorter wavelength
emission in the central regions, which is inconsistent with the overall
trends for the 15$\,\mu$m continuum and [\ion{Ar}{ii}] 6.99$\,\mu$m line.

The nature of the short-wavelength continuum in normal and pure
starburst galaxies is still debated.  It is much weaker than in
systems hosting an active galactic nucleus (AGN) where it is 
attributed to hot dust ($150 - 1700~\mathrm{K}$) close to the AGN
(Genzel \& Cesarsky \cite{Gen00} and references therein).  In non-AGN
galaxies, the existence of $3 - 5~\mathrm{\mu m}$ featureless continuum
from small transiently heated dust grains underlying the PAH features
has been proposed (e.g. Helou et al.~\cite{Hel00}).  On the
other hand, the PAH features are best fitted with Lorentzian profiles
whose wide wings can account for the apparent continuum pedestal
(e.g. Boulanger et al.~\cite{Bou98b}).  Our narrow-band 5.5$\,\mu$m
continuum map of \object{M\,82} provides only tentative evidence of a
more extended distribution than for the 15$\,\mu$m continuum emission
which would suggest a closer association of the emitting particles
with PAHs than VSGs (see also Mattila et al.~\cite{Mat99}).

\section{PAH and continuum emission: relations to the ISM conditions
         and star-forming activity}
        \label{Sect-Spec_distr}

The two most striking aspects seen in the spectra presented in
Fig.~\ref{fig-spec_stbs} are their remarkable similarity in the
$\lambda = 5 - 11~\mathrm{\mu m}$ regime dominated by PAH emission,
and the spread in relative intensity of the longer wavelength
emission attributed to VSGs.  In this section, we focus on these
two spectral components after discussing extinction effects at
MIR wavelengths.

\subsection{The effects of extinction at mid-infrared wavelengths}
            \label{Sub-ext_effects}

Interstellar extinction can significantly affect 
the MIR spectral energy distribution (SED) in obscured sources.
In this regard, Rigopoulou et al.~(\cite{Rig99}) noted the gradual
suppression of the PAH 8.6$\,\mu$m feature in progressively obscured 
starbursts and ULIRGs.  They further proposed that variations in the 
PAH~6.2$\,\mu$m/7.7$\,\mu$m ratio are dominated by extinction effects on
the basis of the trend of decreasing ratio with increasing extinction
observed in a subset of their sample for which reliable extinction
determinations were available.

Another important consideration concerns the trough centered near 9.7$\,\mu$m
traditionally attributed to absorption by interstellar silicate dust grains.
Recent studies have cast doubt on the reliability of extinction estimates
based on the observed feature depth.  Sturm et al.~(\cite{Stu00}) discussed
this issue in detail based on SWS data of \object{M\,82} and \object{NGC\,253},
from a comparison with the spectrum of the Galactic reflection nebula 
\object{NGC\,7023} and from considerations of the relative optical depths
expected for silicate absorption near 9.7 and 18$\,\mu$m.  In particular,
they showed that the \object{M\,82} SWS spectrum, including the dip at
9.7$\,\mu$m, can be well reproduced by a combination of the
\object{NGC\,7023} spectrum
and a power law rising from 8.5$\,\mu$m longwards without invoking large
extinctions.  More generally, the overall spectral invariance in the
$5 - 11~\mathrm{\mu m}$ region among a variety of Galactic and extragalactic
sources which are known or expected to cover a large range in extinction
seems to argue in favour of the 9.7$\,\mu$m ``absorption feature'' being
predominantly due to the gap between the main PAH emission complexes 
(e.g. Helou et al.~\cite{Hel00}).

These suggestions raise the important question of how such an
interpretation can be reconciled with the large and variable
extinction in \object{M\,82}, \object{NGC\,253}, and \object{NGC\,1808}
determined by alternative methods based, for instance, on the
relative intensities of H recombination lines.  To address this
issue, we explored quantitatively the effects of extinction
at MIR wavelengths by applying a range of extinction to
a representative template spectrum constructed in a similar way
as Sturm et al.~(\cite{Stu00}).  We added to a PDR component
(from the ISOCAM data of \object{NGC\,7023}) a power law 
$f_{\nu} \propto (\lambda - 8.5)^{\alpha}$ which represents well
the VSG emission seen in our data.  We adopted an index $\alpha = 1.5$
and scaled the two components so as to obtain resulting spectra similar to
that of \object{M\,82}.  For the purpose of illustrating the effects of
different levels of extinction, the choice of these parameters is irrelevant.

We considered two representative geometries: a uniform foreground dust
screen and a homogeneous mixture of dust and sources.  For these models, the
observed intensity $I({\lambda})$ of the emerging radiation is proportional
to the total intrinsic intensity of the sources $I_{0}(\lambda)$ by the
factors $e^{-\tau_{\lambda}}$ and $(1 - e^{-\tau_{\lambda}})/\tau_{\lambda}$,
respectively, where $\tau_{\lambda}$ is the optical depth related
to the extinction in magnitude via $A_{\lambda} = 1.086\,\tau_{\lambda}$.
We adopted the extinction law proposed by Draine (\cite{Dra89})
and investigated the effects of deviations from this law in the
$3 - 10~\mathrm{\mu m}$ region as found towards the Galactic Center
(Lutz \cite{Lut99}; hereafter simply ``GC law'').  Such deviations
are consistent with the H recombination line spectrum observed with
the SWS in \object{M\,82} (F\"orster Schreiber et al.~\cite{FS01}).
We varied the level of extinction specified in visual magnitudes $A_{V}$
for the mixed model in the range $A_{V}^\mathrm{MIX} = 0 - 700~\mathrm{mag}$.
For a meaningful comparison, we varied $A_{V}$ for uniform foreground
obscuration in the range $A_{V}^\mathrm{UFS} = 0 - 50~\mathrm{mag}$,
which results in the same attenuation at 10$\,\mu$m as the range
considered for the mixed model.  We normalized each extincted
model spectrum to the integrated flux between 6.0 and 6.6$\,\mu$m,
as for the spectra of Fig.~\ref{fig-spec_stbs}.

\begin{figure*}[!ht]
\centering
\resizebox{0.75\hsize}{!}
          {\includegraphics[bb=17 50 578 760,clip]{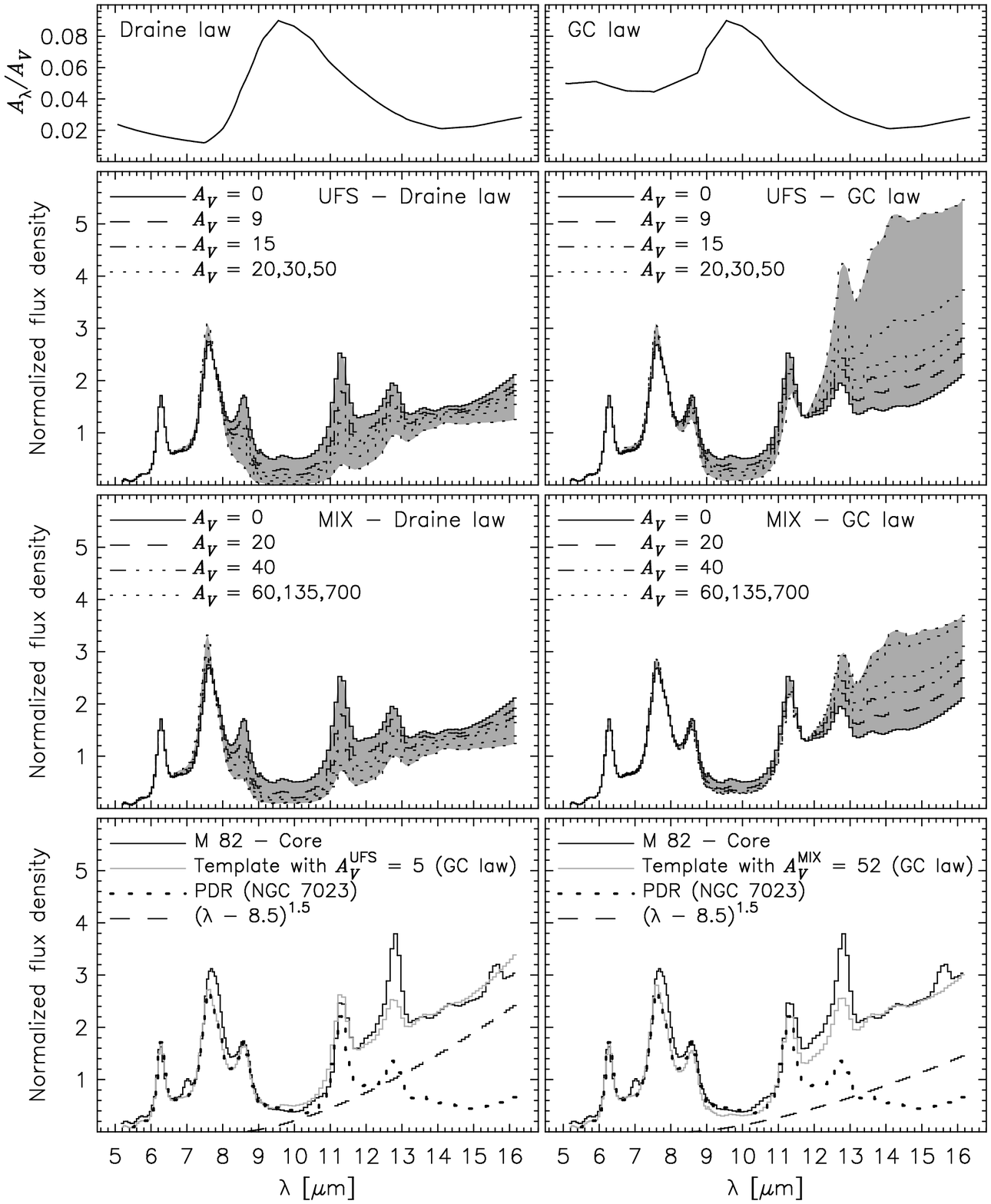}}
\caption
{
Extinction effects at MIR wavelengths.
The top panels show the extinction law from Draine (\cite{Dra89}), and
the modifications to this law at $\lambda \leq 10~\mathrm{\mu m}$
appropriate for the Galactic Center line of sight
(Lutz \cite{Lut99}).
The middle panels show the variations of the SED
of a template spectrum affected by various levels of
extinction, for different dust and source geometries and extinction
laws, as labeled in each plot.  The template 
combines the spectrum of the Galactic PDR NGC\,7023 and
a power-law spectrum (shown individually in the bottom panels
by the dashed and dotted lines, respectively).
``UFS'' and ``MIX'' stand for uniform foreground screen and mixed model.
The grey shading outlines the range of variations for 
$A_{V}^\mathrm{UFS} = 0 - 50~\mathrm{mag}$ and 
$A_{V}^\mathrm{MIX} = 0 - 700~\mathrm{mag}$.
The various curves show the results for
selected extinction levels, corresponding to similar attenuation factors at
$10~\mathrm{\mu m}$ between the two extinction models (labeled in each plot).
The bottom panels show non-formal fits (grey solid line) to the 
spectrum of the starburst core of M\,82 (black solid line), using the
template PDR $+$ power-law spectrum, attenuated by the best-fit
extinction for each geometry assuming the GC extinction law and derived
from radio to optical H recombination line measurements
(see Sect.~\ref{Sub-ext_effects}).
All spectra are normalized to the total flux between 6.0 and
$6.6~\mathrm{\mu m}$, except for the power-law.
}
\label{fig-ext_effects}
\end{figure*}

Figure~\ref{fig-ext_effects} presents our simulations.
The plots outline the distinct behaviours for the two geometries
considered with increasing $A_{V}$.  They especially emphasize
the fact that for the mixed model, differential extinction effects over
the spectrum rapidly reach an asymptotic regime where large variations
in $A_{V}$ become less and less perceptible as $A_{V}$ increases.
In addition, except for the most extreme cases ($A_{V} > 20~\mathrm{mag}$),
the choice of dust and source geometry has little impact.
This is because the absolute levels of extinction are fairly small
and the wavelength dependence of extinction is relatively weak;
the wavelength range covered is thus too limited to probe
appreciably different optical depths which could allow the
discrimination between different dust and sources distributions.

For the Draine extinction law,
the effects are largest between 8 and 13$\,\mu$m.
Our models illustrate well how the PAH 8.6 and 11.3$\,\mu$m features are
substantially suppressed as extinction increases, indicating that their
ratio with the PAH features at 6.2 and 7.7$\,\mu$m can be much affected by
obscuration.  Moreover, for a continuum level defined over the limited
$8 - 13~\mathrm{\mu m}$ interval, relatively small optical depths would be
inferred even if the extinction is in fact large.  Assuming the GC law at
$3 - 10~\mathrm{\mu m}$ results in a remarkably different behaviour.  The
extinction effects are significantly smaller in the $8 - 13~\mathrm{\mu m}$
region and the suppression of the PAH 8.6 and 11.3$\,\mu$m features is much
less important.  The largest effects are observed in the relative level
between the short and long wavelength regions.  This is a consequence
of the much flatter GC law between 3 and 10$\,\mu$m.

Our simulations indicate that dust obscuration adds a significant degree
of degeneracy in the interpretation of the observed MIR emission,
complicating the determination of intrinsic properties.  Depending 
on the extinction law used, the shape of the PAH 8.6 and 11.3$\,\mu$m
features as well as their flux ratio with the PAH 6.2 and 7.7$\,\mu$m
features vary considerably.  Therefore, extinction estimates
based on diagnostics involving the PAH 8.6 or 11.3$\,\mu$m feature may be
strongly biased by the choice of extinction law.  Furthermore, part of the
spread in intensity observed in the long-wavelength continuum relative
to the shorter wavelength emission could be attributed to varying
extinction levels (see Fig.~\ref{fig-spec_stbs} and Sect.~\ref{Sub-cont}
below).

The interpretation of the 9.7$\,\mu$m trough as largely due to the
gap between the flanking PAH complexes appears to hold over a wide range
of extinction, especially for the GC extinction law.  The near-invariance
of the $5 - 11~\mathrm{\mu m}$ spectrum is therefore not inconsistent with
large and/or variable obscuration among and within galaxies.  We illustrate
this with the case of \object{M\,82}, for which a mixed model with
$A_{V} = 52~\mathrm{mag}$ best reproduces the observed H recombination
lines from radio to optical wavelengths while the best-fit uniform 
foreground screen extinction of $A_{V} = 5~\mathrm{mag}$ provides a much
poorer fit (F\"orster Schreiber et al.~\cite{FS01}).  We applied these
two extinction models to the same template as for the simulations
described above.  The resulting spectra are plotted along with the observed
spectrum of \object{M\,82} in the bottom panels of Fig.~\ref{fig-ext_effects}.
We did not attempt to fine-tune the models by formal fitting.
Given the uncertainties on the exact nature of the emitting particles, 
specific assumptions on model parameters are not well constrained and a
simple empirical approach is sufficient for our purposes.  The comparison
shows that strictly from the point of view of the $5 - 16~\mathrm{\mu m}$
range, both extinction models reproduce equally well the observed SED of
\object{M\,82}, thus demonstrating that a high extinction cannot be
excluded from the overall shape of the MIR spectrum alone.
 
We wish to emphasize that we do not dismiss the possibility of silicate
absorption around 9.7$\,\mu$m in general but merely want to point out the
difficulties involved in the interpretation of the observed feature.
The SEDs of moderately to highly obscured sources including ULIRGs
often exhibit a strong dip near 9.7$\,\mu$m together with suppressed
PAH 8.6 and 11.3$\,\mu$m features relative to those at 6.2 and 7.7$\,\mu$m,
consistent with the presence of silicate grains
(e.g. Dudley \& Wynn-Williams \cite{Dud97}; Dudley \cite{Dud99};
Laurent et al.~\cite{Lau00}; Le Floc'h et al.~\cite{LeF02}).
Given the near invariance of the PAH spectrum in a wide range of environments,
it should be possible to define an indicator measuring the differential
extinction between the 9.7$\,\mu$m region and adjacent less affected intervals
that quantifies the absolute extinction.  One must however be aware of the
importance of sufficient wavelength coverage and resolution to assess
properly the impact of the PAH complexes which can make the silicate
absorption look artificially deep, of the dependence of the inferred $A_{V}$
on the assumed extinction law and spectral intervals used to measure the
feature depth, and of the limited sensitivity of this diagnostic leading
to rather large uncertainties in the derived $A_{V}$.
The upcoming launch of SIRTF in 2003 will provide $5 - 40~\mathrm{\mu m}$
spectroscopy with an increase in sensitivity by two orders of magnitude
and will help us address this issue by enabling a better sampling of the
continuum emission as well as both the 9.7 and 18$\,\mu$m silicate bands.

\subsection{The $\mathit{\lambda \geq 11~\mu m}$ continuum emission}
            \label{Sub-cont}

The $\lambda \ga 11~\mathrm{\mu m}$ continuum emission in our data exhibits
a large spread in intensity relative to the shorter wavelength emission.
Similar variations have been seen in normal spirals and starburst
galaxies observed with {\em ISO\/} instruments and, based on
evidence provided by Galactic sources, are generally interpreted in
terms of the relative contribution of \ion{H}{ii} regions to the MIR
emission (e.g. Laurent et al.~\cite{Lau00}; Roussel et al.~\cite{Rou01a};
Dale et al.~\cite{Dal01}).  A close link between the
15$\,\mu$m/7$\,\mu$m colour as measured through the ISOCAM broad-band
LW3 and LW2 filters ($\lambda = 12 - 15~\mathrm{\mu m}$ and
$\rm 5 - 8.5~\mathrm{\mu m}$, respectively) and the {\em IRAS\/}
25$\,\mu$m/12$\,\mu$m colour has been emphasized by
Dale et al.~(\cite{Dal01}) and Roussel et al.~(\cite{Rou01a}).

In order to assess quantitatively the relationship between the 
long-wavelength continuum properties and the \ion{H}{ii} regions,
we investigated the relationship between the 15$\,\mu$m narrow-band
continuum and the [\ion{Ar}{ii}] 6.99$\,\mu$m line emission.
The [\ion{Ar}{ii}] 6.99$\,\mu$m is the most direct probe of
\ion{H}{ii} regions and the least contaminated by PAH emission
available from our ISOCAM data sets.  The \element[0][][][]{Ar}
ionization potential of 15.8~eV is close to that of H (13.6~eV).  Argon
being a noble element, it is, like neon, not expected to be significantly
depleted onto dust grains; while the argon gas-phase abundance may increase
with time as a result of star formation activity and differ among galaxies,
it is not likely to vary significantly over scales of
$\sim 100 - 1000~\mathrm{pc}$ as covered by the starburst regions in our
sample.  For the relatively low nebular excitation in all three sources
(Fig.~\ref{fig-spec_sws}; see also Sturm et al.~\cite{Stu00};
F\"orster Schreiber et al.~\cite{FS01}), argon is mostly singly-ionized so
that the [\ion{Ar}{ii}] 6.99$\,\mu$m emission should trace the \ion{H}{ii}
regions in much the same way as H recombination lines and its luminosity,
scale with the radiation field intensity of the young stars. 
In addition, since the gas-phase abundances of the galaxies are solar
within a factor of a few (e.g. Webster \& Smith \cite{Web83};
Forbes et al.~\cite{For92}; F\"orster Schreiber et al.~\cite{FS01}),
the proportionality factor should be roughly similar.

\begin{figure}
\centering
\resizebox{1.75\hsize}{!}
          {\includegraphics[bb=50 420 587 710,clip]{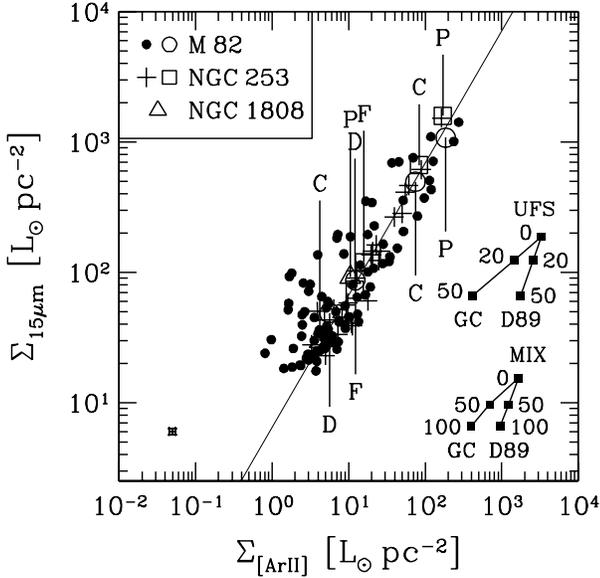}}
\caption
{
Relationship between MIR tracers of H\,{\small II} regions in 
M\,82, NGC\,253, and NGC\,1808.  The diagram shows the variations
of the $15~\mathrm{\mu m}$ continuum luminosity as a function of
the [Ar\,{\small II}] $6.99~\mathrm{\mu m}$ line luminosity normalized
per unit projected area.  Data are plotted for selected regions
(open symbols labeled ``F'' for the ISOCAM field of view,
``D'' for the disk regions, ``C'' for the starburst core,
and ``P'' for the MIR peak).   The filled circles and crosses
show the values for individual resolution elements for M\,82 and
NGC\,253, respectively, within radii of $30^{\prime\prime}$ and 
$15^{\prime\prime}$ (outer disk annulus; table~\ref{tab-apertures});
measurements at $< 3\sigma$ are excluded
(4 points for M\,82, 11 for NGC\,253).
Typical formal uncertainties are indicated by the error bar at the
bottom left and are smaller than the symbol sizes;
they are 5\% for $\Sigma_\mathrm{15\,\mu m}$ and 10\% for
$\Sigma_\mathrm{[Ar\,II]}$ (the median and average differ by $< 1.5\%$).
The straight line is a least-squares fit to the resolution elements'
data accounting for the individual uncertainties.  The effects 
of extinction are shown in the right part of the plot for a
uniform foreground screen and a mixed model (``UFS'' and ``MIX''),
for the extinction law of Draine (\cite{Dra89}, ``D89'') and with
modifications at $3 - 10~\mathrm{\mu m}$ as found towards the Galactic
Center (Lutz \cite{Lut99}, ``GC'').  Selected values of extinction
in visual magnitudes $A_{V}$ are labeled.
}
\label{fig-cont_stbs}
\end{figure}

Figure~\ref{fig-cont_stbs} presents the 15$\,\mu$m continuum versus
[\ion{Ar}{ii}] 6.99$\,\mu$m luminosities for our galaxies, normalized
to unit projected surface area ($\Sigma_\mathrm{15\,\mu m}$ and
$\Sigma_\mathrm{[Ar\,II]}$).  The data are shown for the selected
regions as well as for the individual resolution elements
enclosed within the radii defining the outer limit of the disk
regions in \object{M\,82} and \object{NGC\,253}.  Resolution elements
with measurements at $< 3\sigma$ were excluded.  The data points
follow a well defined distribution, with a remarkable overlap for
the different galaxies.  Least-squares fitting to the resolution
elements' data accounting for the individual formal uncertainties
yields the relation
\begin{equation}
\log(\Sigma_\mathrm{15\,\mu m}) = 1.01\,\log(\Sigma_\mathrm{[Ar\,II]})
+ 0.807,
\label{Eq-sfr}
\end{equation}
with dispersion around the fit of 0.203~dex.  The extinction paths
in the diagram of Fig.~\ref{fig-cont_stbs} are nearly parallel to
the locus formed by the data points; however, implausibly large 
differential extinctions would be required in order to account
entirely for the observed distribution.

The essentially linear proportionality we find between 
$\Sigma_\mathrm{15\,\mu m}$ and $\Sigma_\mathrm{[Ar\,II]}$
indicates that the 15$\,\mu$m emission provides a good quantitative
indicator of the star-forming activity in starburst environments,
to within uncertainties determined by the dispersion of the data
and by extinction (a factor of 2.5 for an $A_{V} = 50~\mathrm{mag}$
assuming purely foreground extinction).  Roussel et al.~(\cite{Rou01b})
reached a similar conclusion for more quiescent spiral disks where
the $12 - 15~\mathrm{\mu m}$ emission scales linearly with the H$\alpha$
line flux.  Part of the scatter in Fig.~\ref{fig-cont_stbs} may result
from variations in the physical conditions and exact composition of the
gas and dust within and between the galaxies.  More important factors,
however, are variations in relative spatial distribution of the emission
and possibly of the excitation state of the gas.

The observed relationship holds remarkably well in view of the differences
in morphology of the 15$\,\mu$m continuum and [\ion{Ar}{ii}] 6.99$\,\mu$m
line emission as seen in Figs.~\ref{fig-M82_maps}, \ref{fig-N253_maps},
\ref{fig-N1808_maps}, and \ref{fig-profiles}.  The correlation extends however
over two to three orders of magnitude, much larger than the dispersion of a
factor of 1.6.  Undoubtedly, the relative spatial variations between both
tracers contribute significantly to the scatter.  Within the starburst cores
of \object{M\,82} and \object{NGC\,253}, where the steeply rising SEDs at
$\lambda \ga 11~\mathrm{\mu m}$ are dominated by VSG emission, the differences
may be attributed to different ranges in exciting photon energies for VSGs
and [\ion{Ar}{ii}] 6.99$\,\mu$m line as well as to extinction effects
(Sect.~\ref{Sect-Spat_distr}).  For \object{NGC\,1808}, the spatial
distributions differ probably because a different dust/particles
population produces the flat long-wavelength continuum
(Sect.~\ref{Sect-Spat_distr}).
Nevertheless, the data for \object{NGC\,1808} and for the disk regions
of \object{M\,82} and \object{NGC\,253}, characterized by flat
$\ga 11~\mathrm{\mu m}$ SEDs, are well described by Eq.~\ref{Eq-sfr}.

Variations of the excitation state of the photoionized nebulae may influence
the $\Sigma_\mathrm{15\,\mu m}$ versus $\Sigma_\mathrm{[Ar\,II]}$ relationship
as well.  In particular, the [\ion{Ar}{iii}] 8.99$\,\mu$m line is fairly strong
in the {\em ISO\/}-SWS spectrum of \object{M\,82} (Fig.~\ref{fig-spec_sws})
but the [\ion{Ar}{iii}] 8.99$\,\mu$m/[\ion{Ar}{ii}] 6.99$\,\mu$m ratio of
0.18 (0.26 after extinction correction) is low and abundance estimates
indicate that $\sim 25\%$ only of the argon is doubly ionized
(F\"orster Schreiber et al.~\cite{FS01}).
As emphasized in Sect.~\ref{Sect-Spat_distr}, the spatial distributions
of various MIR fine-structure lines and Br$\alpha$ are similar and suggest
a roughly constant excitation state of the \ion{H}{ii} regions across
\object{M\,82} (Achtermann \& Lacy \cite{Ach95}).  This is further confirmed
by the nearly uniform \ion{He}{i} 2.06$\,\mu$m/Br$\gamma$ ratio in the
central $260 \times 160~\mathrm{pc}$ which corresponds closely to
the excitation derived from the SWS
[\ion{Ne}{iii}] 15.56$\,\mu$m/[\ion{Ne}{ii}] 12.81$\,\mu$m,
[\ion{Ar}{iii}] 8.99$\,\mu$m/[\ion{Ar}{ii}] 6.99$\,\mu$m, and
[\ion{S}{iv}] 10.5$\,\mu$m/[\ion{S}{iii}] 18.7$\,\mu$m ratios
within larger apertures up to $430 \times 225~\mathrm{pc}$
(F\"orster Schreiber et al.~\cite{FS01}).
Similar arguments are more difficult for \object{NGC\,253} and
\object{NGC\,1808} because of the lack of relevant data.  However, the
[\ion{Ar}{iii}] 8.99$\,\mu$m/[\ion{Ar}{ii}] 6.99$\,\mu$m $\approx 0.03$
measured from the SWS spectrum of \object{NGC\,253} is
substantially lower than for \object{M\,82}, as is the
[\ion{Ne}{iii}] 15.56$\,\mu$m/[\ion{Ne}{ii}] 12.81$\,\mu$m ratio
(0.06 compared to 0.18; Thornley et al.~\cite{Tho00}; see also
Giveon et al.~\cite{Giv02} for the relationship between these two MIR line
ratios).  For \object{NGC\,1808}, we can only note from the ISOCAM spectra
(Fig.~\ref{fig-spec_stbs}) that the weakness of the 15.7$\,\mu$m feature
and the non-detection of [\ion{Ar}{iii}]~8.99$\,\mu$m indicate low
excitation supporting that argon mostly is singly ionized.
Summarizing, the excitation state of the gas might cause for \object{M\,82}
a general but small offset to the left in the relationship of
Fig.~\ref{fig-cont_stbs} compared to the other galaxies because of the larger
fraction of $\mathrm{Ar^{++}}$ while spatial variations are not likely to
introduce scatter larger than this offset.  These effects are expected to be
smaller for both \object{NGC\,253} and \object{NGC\,1808}.

The relationship between the 15$\,\mu$m continuum and star formation intensity
is supported by the analysis of a larger sample including spiral and starburst
galaxies that will be presented in a forthcoming paper (F\"orster Schreiber,
Roussel, \& Sauvage, in prep.).  In this paper, based on the previous work of
Roussel et al.~(\cite{Rou01b}) for spiral disks, we find that the 15$\,\mu$m
continuum as well as the 7$\,\mu$m PAH-dominated emission correlate well with
the star formation rate over nearly 7 orders of magnitude, with the interesting
distinction of two regimes that correspond to quiescent star formation in disks
and more intense activity in circumnuclear regions and starbursts.

\subsection{The $\mathit{\lambda = 5 - 11~\mu m}$ emission}  
            \label{Sub-invar}

The near invariance of the $5 - 11~\mathrm{\mu m}$ spectrum
seen in our data of \object{M\,82}, \object{NGC\,253}, and
\object{NGC\,1808} has been noted in a number of other studies of different
types of galaxies powered mainly by star formation, as well as in a variety
of Galactic sources (see Tielens et al.~\cite{Tie99} for a review;
see also e.g. Boulanger et al.~\cite{Bou98a}; Helou et al.~\cite{Hel00};
Uchida et al.~\cite{Uch00}).  This indicates that the PAHs are very stable
under a wide range of physical conditions despite their small sizes
(typically $\sim 100$ atoms) and is taken as direct observational
evidence for the stochastic nature of the emission processes involved.
Furthermore, if different types of PAHs coexist, the near
constancy in the $5 - 11~\mathrm{\mu m}$ region suggests that their
relative abundances vary little.  Finally, it has consequences on the
interpretation of and extinction measurements from the 9.7$\,\mu$m
silicate absorption feature because of the intrinsic gap between the
main $6 - 9~\mathrm{\mu m}$ and $11 - 13~\mathrm{\mu m}$ PAH complexes
(Sect.~\ref{Sub-ext_effects}).

Variations of the relative intensities of the PAH features do exist, however.
We focus on the PAH~6.2$\,\mu$m/7.7$\,\mu$m and PAH~8.6$\,\mu$m/11.3$\,\mu$m
ratios, maps of which were presented for \object{M\,82} and 
\object{NGC\,253} in Sect.~\ref{Sub-images}.  Of particular interest is also
the enhancement of the PAH 11.3$\,\mu$m relative to the other features in
\object{NGC\,1808} compared to \object{M\,82} and \object{NGC\,253}, evident
in Fig.~\ref{fig-spec_stbs}.  The variations of PAH ratios in our sample
are comparable to those observed within and between Galactic sources
of similar types (e.g. Cesarsky et al.~\cite{Ces96a};
Boulanger et al.~\cite{Bou98a}; Lu \cite{Lu98};
Cr\'et\'e et al.~\cite{Cre99}; Uchida et al.~\cite{Uch00}), along the disk
of the spiral galaxy \object{NGC\,891} (Mattila et al.~\cite{Mat99}), and
among a sample of 15 starbursts and ULIRGs (Rigopoulou et al.~\cite{Rig99}).
A detailed discussion of PAH ratios is beyond the scope of this paper, but
we briefly mention possible interpretations of the variations seen in our
data in the light of some recent theoretical and empirical work.

\begin{figure*}[!ht]
\includegraphics[bb=25 210 590 620,width=12.0cm,clip]{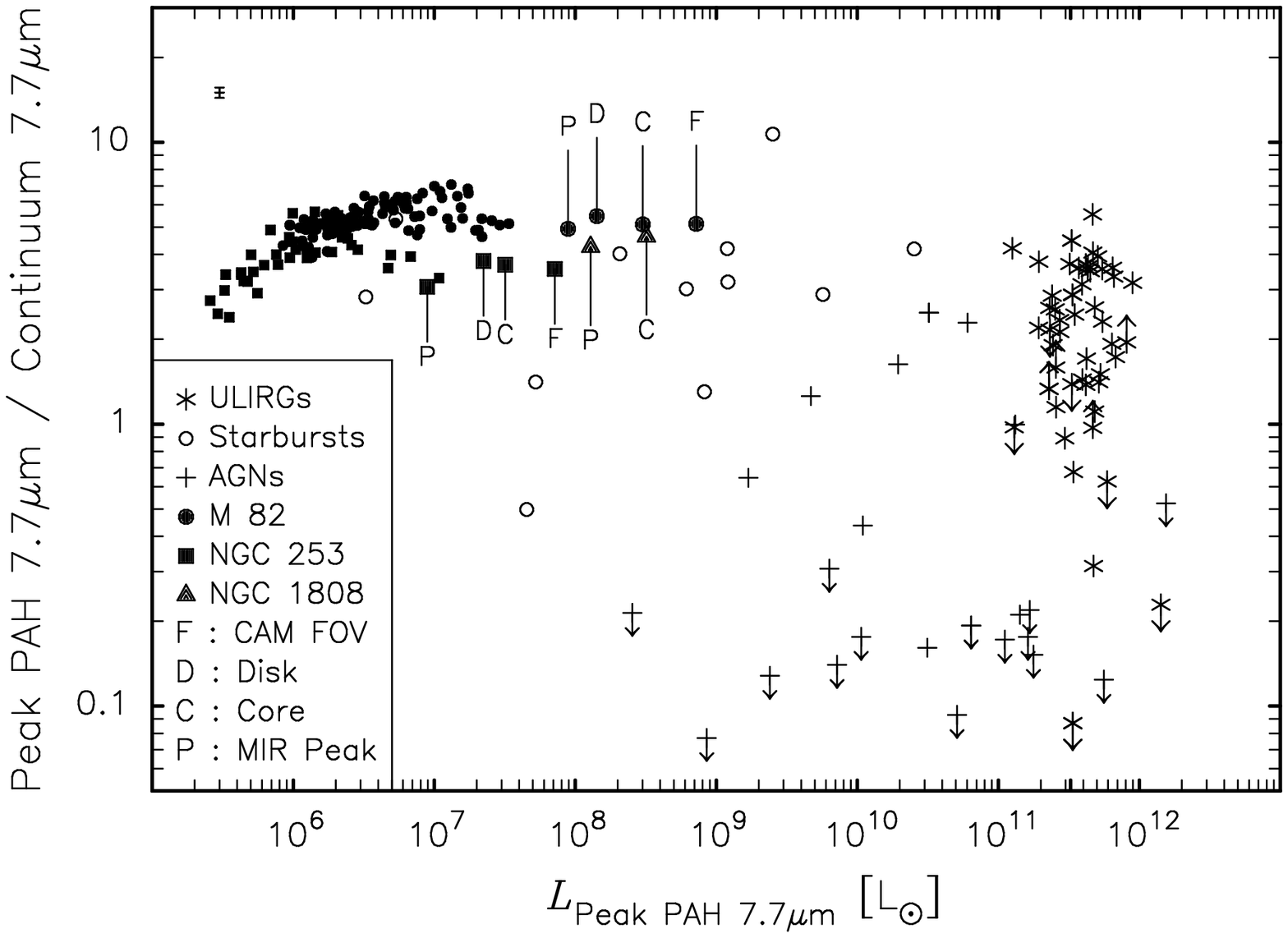}
\caption
{
Diagnostic diagram for dominant starburst versus AGN activity.
The data for selected regions in M\,82, NGC\,253, and NGC\,1808
as well as individual resolution elements for M\,82 and NGC\,253
(large labeled and small unlabeled filled symbols, respectively;
see inset) are compared to the global properties for the samples of
starburst galaxies, AGNs, and ULIRGs of Rigopoulou et al.~(\cite{Rig99};
open circles, crosses, and stars).  For M\,82 and NGC\,253,
resolution elements within radii of $30^{\prime\prime}$ and
$15^{\prime\prime}$ (outer disk annulus; table~\ref{tab-apertures}) are
plotted, and typical formal uncertainties are indicated by the error bar
at the top left: $< 1\%$ for $L_\mathrm{Peak~PAH\,7.7\mu m}$ and 4\% for 
PAH 7.7$\,\mu$m L/C ratio (uncertainties are comparable to or smaller than
the symbol sizes; average and median values differ by $< 1\%$).
}
\label{fig-peak_pixels}
\end{figure*}

Since the emission mechanism is stochastic in nature, PAH ratios
are not directly related to the SED of the incident radiation
(e.g. Boulanger et al.~\cite{Bou98a}; Uchida et al.~\cite{Uch00}).
On the other hand, they may carry indirect information since PAHs
exposed to hard and intense radiation fields can be ionized, lose
H atoms, or be photodissociated (e.g. L\'eger et al.~\cite{Leg89};
Allamandola et al.~\cite{All89}; Schutte et al.~\cite{Sch93};
Allamandola et al.~\cite{All99}; Hudgins \& Allamandola \cite{Hud99}).
In particular, the PAH~8.6$\,\mu$m/11.3$\,\mu$m
is believed to trace the fraction of singly ionized
to neutral PAHs, presumably driven by the strength of the radiation field
from the OB stars (see references above, and Joblin et al.~\cite{Job96}).
Draine \& Li (\cite{Dra01}) presented a thorough calculation of the expected
PAH spectrum as a function of various parameters including PAH size, charging
conditions, and starlight intensity.  Their results indicate that the
PAH~6.2$\,\mu$m/7.7$\,\mu$m depends primarily on PAH size while the
PAH~11.3$\,\mu$m/7.7$\,\mu$m is mainly sensitive to the fraction of ionized
versus neutral PAHs and only modestly to PAH size.  The effects of radiation
field become noticeable only at high intensities ($\ga 10^{5}$ the
average local Galactic far-UV flux in Habing units) and for large PAHs with
$\ga 10^{2}$ carbon atoms.  Mattila et al.~(\cite{Mat99}) suggested that
PAH~6.2$\,\mu$m/7.7$\,\mu$m variations may be due to differences in average
temperature of the PAHs during their temperature spikes, related to PAH size
or mean exciting UV photon energies.  Alternatively, they could be attributed
to broadening of the 7.7$\,\mu$m feature at low radiation field energy 
densities (Uchida et al.~\cite{Uch00}); this would reduce the PAH 7.7$\,\mu$m
flux and increase the continuum level within our fixed bandpasses.
In addition to intrinsic variations, extinction can significantly alter
the shape and relative intensities of the PAH features as shown in
Sect.~\ref{Sub-ext_effects}.

Of our sample galaxies, \object{M\,82} offers the most interesting case for
interpretation.  The overall bilobal structure in the maps of PAH ratios 
and CO gas distribution (Fig.~\ref{fig-M82_CAM_CO}) might result from the
different composition and physical processes that the emitting PAHs undergo
when exposed to the varying conditions across \object{M\,82}, from the
starburst core to the more quiescent disk via the transition regions marked
by the molecular gas ring.  The maxima in PAH~8.6$\,\mu$m/11.3$\,\mu$m
ratio lie at smaller radii than the minima in PAH~6.2$\,\mu$m/7.7$\,\mu$m
ratio, possibly indicating a higher degree of PAH ionization within the
most intense starburst sites at the inner edge of the molecular ring;
this is particularly striking southwest of the nucleus where the peak
PAH~8.6$\,\mu$m/11.3$\,\mu$m coincides very well with the location of
the most prominent \ion{H}{ii} region complexes (Sect.~\ref{Sub-M82images}).
The variations of PAH~6.2$\,\mu$m/7.7$\,\mu$m ratio could reflect differences
in the PAH size distribution, combined with extinction effects, where larger
PAHs can better form and survive in denser, more shielded environments
associated with molecular gas concentrations.
In \object{NGC\,253}, the peak PAH~8.6$\,\mu$m/11.3$\,\mu$m ratio at the
nucleus where the starburst is mainly occurring could also be due to a
larger fraction of ionized PAHs.  Finally, the enhanced PAH 11.3$\,\mu$m
feature in \object{NGC\,1808} compared to \object{M\,82} and
\object{NGC\,253} could be explained by a more neutral 
mixture of PAHs and a more diffuse radiation field.

The relative intensity of PAH and continuum emission can also hold information
on the physical environment within astronomical sources.  In particular, the
ratio of the peak intensity of the PAH 7.7$\,\mu$m to the underlying continuum,
hereafter PAH 7.7$\,\mu$m L/C ratio, has been shown to constitute a powerful
discriminator between star formation activity or an AGN
as the main source of the bulk of infrared luminosity
(Rigopoulou et al.~\cite{Rig99}; see also Genzel et al.~\cite{Gen98};
Laurent et al.~\cite{Lau00}; Tran et al.~\cite{Tra01}).
Starburst-dominated objects are characterized by ratios $\ga 1$ while
AGN-dominated ones have ratios $\la 1$.  We compared measurements of
the PAH 7.7$\,\mu$m L/C ratio from our data of
\object{M\,82}, \object{NGC\,253}, and \object{NGC\,1808} with those
obtained by Rigopoulou et al.~(\cite{Rig99}) for a large sample including
starbursts, AGNs, and ULIRGs.  Our motivation was to assess how much this
indicator depends on the source luminosity at the faint end and compare the
spread in global ratios among different galaxies with the spatial variations 
within individual galaxies.

Figure~\ref{fig-peak_pixels} shows the PAH 7.7$\,\mu$m L/C ratio versus
peak PAH 7.7$\,\mu$m luminosity for selected regions of our galaxies,
for individual resolution elements of \object{M\,82} and \object{NGC\,253}
(within the radii defining the outer 
limit of the disk regions in table~\ref{tab-apertures}), and for the
Rigopoulou et al.~(\cite{Rig99}) sample.  We measured the PAH 7.7$\,\mu$m
peak intensity and underlying continuum according to the definitions of
Rigopoulou et al.  We note that with these definitions, the continuum
might be underestimated in highly obscured sources as a result of
extinction affecting notably the 11$\,\mu$m region (as also discussed
by Laurent et al.~\cite{Lau00}).  We chose the PAH 7.7$\,\mu$m peak
luminosity as other characteristic property, taken as an approximate
indicator of the total infrared luminosity.  Although the PAH to infrared
luminosity ratio can vary by up to factors of several in different
environments (Rigopoulou et al.~\cite{Rig99}), our assumption has little
consequences on the interpretation of Fig.~\ref{fig-peak_pixels} since the
data span a range in luminosity extending over more than six orders of
magnitude\footnote{Additional though indirect support for our assumption
is provided by the relationship between the 7$\,\mu$m flux measured through
the ISOCAM LW2 filter ($5 - 8.5~\mathrm{\mu m}$) and the far-infrared (FIR,
$43 - 123~\mathrm{\mu m}$) flux obtained by Roussel et al.~(\cite{Rou01b})
for galactic disks, with $F_\mathrm{7\,\mu m} \propto (F_\mathrm{FIR})^{1.2}$.
Furthermore, Dale et al.~(\cite{Dal01}) derived a typical value of
$\approx 15\%$ for the $5 - 20~\mathrm{\mu m}$ to FIR luminosity ratio in
normal galaxies while Charmandaris et al.~(\cite{Cha02}) found this ratio
to be $\sim 5\%$ for three ULIRGs.}.

Our data of \object{M\,82}, \object{NGC\,253},
and \object{NGC\,1808} extend very well the trend defined by the
global properties of pure starburst galaxies and ULIRGs, populating the
PAH 7.7$\,\mu$m L/C $\ga 1$ region down to luminosities about an order
of magnitude lower.  The ratios in all three galaxies lie well above the
starburst--AGN separation at a ratio of unity and form a tighter distribution.
The average and $1\sigma$ dispersion for the individual resolution elements
in \object{M\,82} and \object{NGC\,253} are $5.0 \pm 0.9$ compared to
$3.7 \pm 2.5$ for the pure starbursts of Rigopoulou et al. and $2.8 \pm 1.0$
for their ULIRGs (excluding those with ratios $\leq 1$ or with only limits on
the measurements).  It is also interesting to note that the starburst trend
holds for regions on spatial scales ranging from $\approx 60~\mathrm{pc}$
for the smallest individual regions in \object{NGC\,253} up to several
kiloparsecs for the largest starbursts and ULIRGs (see the
near-infrared images of Rigopoulou et al.~\cite{Rig99}).

\section{Summary and conclusions}  \label{Sect-conclu}

We have presented ISOCAM $\lambda = 5.0 - 16.5~\mathrm{\mu m}$
spectrophotometric imaging of the starburst galaxies \object{M\,82},
\object{NGC\,253}, and \object{NGC\,1808}.  The spectrum of all three
objects, down to the smallest scales of $\sim 100~\mathrm{pc}$ accessible
from our data, exhibit similar characteristics including prominent PAH bands,
a featureless continuum most obvious at $\lambda \ga 11~\mathrm{\mu m}$, and
a trough in the 10$\,\mu$m region.
We securely identified the main emission features detected
in the $R \sim 40$ ISOCAM data of \object{M\,82} and \object{NGC\,253}
based on their $R \sim 500 - 1000$ SWS spectra.
The comparison emphasizes the caution that should be exercised when
interpreting low resolution data because of potential blends or
misidentifications between emission lines from ionized gas and PAH features
originating in PDRs, a notable example being [\ion{Ne}{ii}] 12.81$\,\mu$m
and PAH 12.7$\,\mu$m.  Using a simple model combining a template PDR spectrum
and a power-law $f_{\nu} \propto (\lambda - 8.5)^{1.5}$, we constructed
a representative starburst SED and explored the effects of extinction at
MIR wavelengths.  Our simulations illustrate the importance of the assumed
extinction law (e.g. the widely used Draine \cite{Dra89} law versus the
Galactic Center law of Lutz \cite{Lut99}) and of the intrinsic PAH
spectrum (especially the gap between the main $6 - 9~\mathrm{\mu m}$
and $11 - 13~\mathrm{\mu m}$ complexes) in shaping the SED of
astronomical sources.  This complicates the interpretation of PAH
ratios as well as extinction measurements relying on the silicate
dust absorption at 9.7$\,\mu$m (see also Sturm et al.~\cite{Stu00}).

As observed previously in a wide range of Galactic and extragalactic sources,
the $5 - 11~\mathrm{\mu m}$ spectrum in our galaxies is nearly invariant.
The relative PAH intensities exhibit nevertheless measureable and significant
variations of $20\% - 100\%$ which may be attributed to various, possibly
interrelated effects including the intensity of the incident radiation field
and the PAH size distribution, ionization, and dehydrogenation.
In our sample, \object{M\,82} probably best illustrates variations of
PAH ratios due to an increased fraction of ionized PAHs within the most
intense starburst sites and, admittedly speculatively, perhaps also to
differences in typical PAH sizes depending on the molecular gas
concentrations.
The PAH~7.7$\,\mu$m L/C ratio in all three galaxies clearly
lies in the range observed for pure starburst systems and extends the trend
reported previously by Rigopoulou et al.~(\cite{Rig99}) to lower luminosities.

In contrast, the $\lambda \ga 11~\mathrm{\mu m}$ region varies most among
our sample galaxies and the ISOCAM maps show a comparatively more compact
15$\,\mu$m continuum distribution relative to the PAH emission.
Strong 15$\,\mu$m continuum in \object{M\,82} and \object{NGC\,253}
indicates an important contribution by VSGs contrary to the case of
\object{NGC\,1808} where the long-wavelength emission is much flatter
presumably because of negligible contribution by VSGs.
We find, however, that in all three galaxies the 15$\,\mu$m
continuum and [\ion{Ar}{ii}] 6.99$\,\mu$m line fluxes satisfy a linear
relationship.  We infer from this that the 15$\,\mu$m continuum provides
a good indicator of star formation activity in starbursts, complementing
the similar results of Roussel et al.~(\cite{Rou01b}) for galactic disks.
In a broader perspective, our galaxies fit well in the trend of increasing
ISOCAM 15$\,\mu$m/7$\,\mu$m ratios with higher levels of star formation
activity found among normal disk galaxies and starburst-powered LIRGs/ULIRGs
(e.g. Laurent et al.~\cite{Lau00}; Roussel et al.~\cite{Rou01b};
Dale et al.~\cite{Dal01}).

The value of the ISOCAM spectrophotometric imaging of
\object{M\,82}, \object{NGC\,253}, and \object{NGC\,1808}
presented in this paper also lies in that it complements existing MIR data
with maps of PAH features, fine-structure lines, and continuum components not
previously imaged.
The poorer angular resolution of ISOCAM compared to that achieved with
large ground-based telescopes is compensated by the larger field of view
revealing more of the large scale emission.  In that respect, we stress the
small size of the MIR source relative to the optical extent of all three
galaxies: the ISOCAM maps cover the central $\approx 1.5~\mathrm{kpc}$
for \object{M\,82}, 0.6~kpc for \object{NGC\,253}, and
5~kpc for \object{NGC\,1808} while the optical diameters are
about 10~kpc for \object{M\,82} and 20~kpc for the other two.
By measuring the flux density within the ISOCAM LW10 filter bandpass,
equivalent to the {\em IRAS\/} 12$\,\mu$m band, we recover all of the
{\em IRAS\/} 12$\,\mu$m emission in the entire field of view for
\object{M\,82} and \object{NGC\,1808}, and about 70\% for \object{NGC\,253}.
This suggests that the total MIR emission is strongly dominated by
the starburst sites in the nuclear regions while the more quiescent star
formation taking place in the disk at larger radii does not contribute much.
It will be interesting to see whether the MIR cameras on board SIRTF,
to be launched in 2003, will confirm this result or discover faint diffuse
emission, especially at shorter ($3 - 5~\mathrm{\mu m}$) wavelengths.

\begin{acknowledgements}

We are grateful to E. Sturm for making the SWS spectrum of
\object{NGC\,253} available to us in electronic form.
We warmly thank H. Roussel, A. Vogler, S. Madden, and especially
D. Tran for many interesting discussions on various aspects of 
this work.  We also wish to thank the referee, Dr.~N. Bergvall,
for his valuable comments that improved the quality of the paper.
VC would like to acknowledge the partial support of
JPL contract 960803.

\end{acknowledgements}



\end{document}